\documentclass{aastex63}
\usepackage{amsmath}
\usepackage{lineno}

\begin{document}

\title{Thermal Inertia Controls on Titan’s Surface Temperature and Planetary Boundary Layer Structure}

\correspondingauthor{Sooman Han}
\email{sooman.han@yale.edu}

\author[0000-0003-1652-5365]{Sooman Han}
\affiliation{Department of Earth and Planetary Sciences, Yale University, New Haven, CT, USA}

\author{Juan M. Lora}
\affiliation{Department of Earth and Planetary Sciences, Yale University, New Haven, CT, USA}



\begin{abstract}
Understanding Titan’s planetary boundary layer (PBL)—the lowest region of the atmosphere influenced by surface conditions—remains challenging due to Titan’s thick atmosphere and limited observations. Previous modeling studies have also produced inconsistent estimates of surface temperature variability, a critical determinant of PBL behavior, often without clear explanations grounded in surface energy balance. Here, we develop a theoretical framework and apply a three-dimensional dry general circulation model (GCM) to investigate how surface thermal inertia influences surface energy balance and temperature variability across diurnal and seasonal timescales. At diurnal timescales, lower thermal inertia surfaces experience larger temperature fluctuations and enhanced daytime sensible heat fluxes due to less efficient subsurface heat conduction. In contrast, at seasonal timescales, surface temperature variations show weak sensitivity to thermal inertia, as atmospheric damping tends to dominate over subsurface conduction. The PBL depth ranges from a few hundred meters to 1,000~m on diurnal timescales, while seasonal maxima reach 2,000~m, supporting the interpretation from a previous study that the Huygens probe captured the two PBL structures. Simulated seasonal winds at the Huygens landing site successfully reproduce key observed features, including near-surface retrograde winds and meridional wind reversals within the lowest few kilometers, consistent with Titan’s cross-equatorial Hadley circulation. Simulations for the planned Dragonfly landing site predict shallower thermal PBLs with smaller fluctuation amplitudes, while maintaining similar wind patterns. This work establishes a physically grounded framework for understanding Titan’s surface temperature and boundary layer variability, and offers a unified explanation of Titan’s PBL behavior that provides improved guidance for future missions.
\end{abstract}



\section{Introduction} 
The planetary boundary layer (PBL) is the lowest atmospheric layer on terrestrial worlds, and is characterized by significant fluxes of heat, moisture, and momentum between the surface and atmosphere, driven by turbulent motions. On Earth, the PBL undergoes substantial diurnal and seasonal variations due to surface heating and convection. These transitions between temperature extrema, whether diurnal or seasonal, generate distinct atmospheric stability regimes that influence critical weather and climate processes, including cloud formation, precipitation, local breezes, and thunderstorms \citep{Stull88, Garratt92, Lee16}. The phenomenon of the PBL extends beyond Earth to other celestial bodies possessing both a solid surface and an atmosphere \citep{Petrosyan11, Flasar14, Lebonnois18}. As such, studying the PBL of other planetary bodies provides valuable insights into general PBL processes, making it crucial for comparative planetology. Mars, for instance, exhibits a PBL similar to Earth's, where convection plays a fundamental role \citep{Hinson08, Spiga10}.

Titan, Saturn's largest satellite, presents unique geological landscapes with dry, vast linear dunes at the equator contrasting with lakes and seas in the polar regions \citep{Lorenz06, Aharonson14, Hayes16}. These striking geographical contrasts make understanding Titan's PBL essential for explaining the formation and evolution of such diverse landscapes. Due to Titan's vast distance from the Sun and its thick, opaque atmosphere, average surface insolation is estimated at approximately 0.1 \% of Earth's, around 1.0~W$\cdot$m$^{-2}$ \citep{McKay89, Mckay91, Tomasko08c}. However, given Titan's extended day ($\sim$16 Earth days) and year ($\sim$29.5 Earth years), the accumulated surface energy over these cycles may suffice to generate notable surface temperature variations and PBL responses. Observational evidence remains limited, especially for a diurnal cycle, with only one in-situ vertical thermal profile near the equatorial surface from the Huygens landing \citep{Fulchignoni05} and scattered radio occultation profiles across the moon \citep{Flasar81, Lindal83, Samuelson97, Courtin02, Schinder11, Schinder12, Schinder20}. Consequently, modeling studies are vital to supplementing the limited observational data and revealing whether Titan's PBL responds to surface heating through convective processes in ways analogous to Earth and Mars, both diurnally and seasonally.

Understanding how surface temperature responds to solar forcing is essential for interpreting variations in the PBL. Within the context of modeling surface temperature variations on terrestrial planets, thermal inertia, defined as $I = \sqrt{\rho c_p k}$ (J$\cdot$m$^{-2}\cdot$s$^{-0.5}\cdot$K$^{-1}$; hereafter we will use thermal inertia units; TIU), serves as a crucial parameter for quantifying a surface's thermal response to energy input \citep{Price77, Putzig05, Mackenzie19b}. Here, $\rho$ represents the density (kg$\cdot$m$^{-3}$), $c_p$ denotes the specific heat capacity (J$\cdot$kg$^{-1}\cdot$K$^{-1}$), and $k$ is the thermal conductivity (W$\cdot$m$^{-1}\cdot$K$^{-1}$) of the surface. Materials with lower thermal inertia tend to exhibit larger temperature variations, reflecting their diminished capacity to store and conduct heat \citep{Stull88, Garratt92}.

Using a dry general circulation model (GCM) of Titan’s atmosphere coupled with a simplified subsurface scheme, \cite{Tokano05} reported diurnal and seasonal surface temperature variations of a few kelvins near the equator—close to the latitude of the Huygens landing at 10.3°S—assuming a uniform thermal inertia of $I = 340$ TIU, characteristic of porous ice regolith. \cite{Williams12} employed a one-dimensional energy-balance model, considering the Huygens landing site with a fixed near-surface atmospheric temperature, to investigate surface temperature variations. Their results indicated smaller diurnal and seasonal surface temperature variations, with amplitudes less than 0.1~K under dry conditions and a thermal inertia of $I = 340$~TIU, contradicting the findings of \cite{Tokano05}. Subsequently, \cite{Charnay12} reported diurnal fluctuations around 0.3~K in surface temperatures at the Huygens landing site using a dry GCM with $I = 400$ TIU \citep{Lebonnois12}. They proposed that the Huygens measurement of a near-adiabatic layer reaching 300~m \citep{Tokano06b} reflected ongoing convection during the probe's descent (at approximately 10 am local time), while changes in the lapse rate at 800~m and 2,000~m corresponded to residual signatures of maximum diurnal PBL extension and seasonal Hadley circulation effects, respectively. More recently, \cite{Mackenzie19b} implemented a global thermal inertia map in a dry GCM \citep{Lora15}, demonstrating that seasonal surface temperature patterns show limited sensitivity to local thermal inertia values.

Previous studies reveal significant inconsistencies in our understanding of Titan’s surface temperature and, consequently, PBL variations. Notably, they report differing magnitudes of diurnal surface temperature variations even when assuming the same or similar values of thermal inertia. Equally puzzling is that seasonal surface temperature variations show little to no sensitivity to thermal inertia within a model, contradicting the expectation that lower thermal inertia surfaces should experience larger temperature swings. The underlying causes of these discrepancies remain unclear. A central limitation of prior work is the absence of a systematic analysis linking surface temperature variations to the governing surface energy balance.

Building on these insights, this study develops a theoretical framework that links surface temperature variations to thermal inertia and surface energy balance, and conducts dry GCM simulations to investigate how varying thermal inertia influences surface temperature and PBL behavior on Titan across diurnal and seasonal timescales. Section~2 introduces the theoretical foundations governing surface temperature variability and outlines the modeling approach. Section~3 presents GCM simulation results for surface temperature variations and demonstrates how the theoretical framework explains them at both diurnal and seasonal timescales. Section~4 examines PBL structures from GCM simulations and their relationship to surface temperatures. Section~5 evaluates expected conditions at the Dragonfly landing site. Section~6 discusses the implications of the findings and concludes by highlighting the distinct roles of atmospheric damping and subsurface heat conduction in regulating surface temperature and PBL structures—dominant at seasonal and diurnal timescales, respectively. Overall, this study provides a critical reassessment of the atmospheric profile measured by Huygens and offers new insights into Titan’s lower atmosphere, informing expectations for the upcoming Dragonfly mission.


\section{Methods}
\subsection{Theory}\label{Theory}

Energy balance at the surface for a dry case (i.e., without latent heat flux) is given by:

\begin{linenomath*}
\begin{equation}
R_\text{net}+H=Q_\text{net}
\end{equation}
\end{linenomath*}

\noindent where $R_\text{net}$ and $H$~(W$\cdot$m$^{-2}$) represent the net radiative flux into and the sensible heat flux out of the surface, respectively. The residual, $Q_\text{net}$~(W$\cdot$m$^{-2}$), is the net energy flux at the surface available for subsurface heat conduction.

The evolution of surface temperature follows the subsurface heat diffusion equation using $Q_\text{net}$ as the surface boundary condition:

\begin{linenomath*}
\begin{equation}
\rho c_p \frac{\partial T(z,t)}{\partial t} = \frac{\partial}{\partial z}\left(k\frac{\partial T(z,t)}{\partial z} \right)
\end{equation}
\end{linenomath*}

\begin{linenomath*}
\begin{equation}
\left. k \frac{\partial T(z,t)}{\partial z} \right|_{z=0} = Q_{\text{net}}
\end{equation}
\end{linenomath*}

\noindent where $\rho c_p$ (J$\cdot$m$^{-3}\cdot$K$^{-1}$) collectively denotes volumetric heat capacity. Here, homogeneous subsurface properties are considered for simplicity and to compare with previous results. The net energy flux at the surface can be simplified as a single harmonic oscillator with linear atmospheric damping \citep[e.g.,][]{Dickinson88}:

\begin{linenomath*}
\begin{equation}
Q_{\text{net}} \approx Q_0\exp(i\omega t) - a(T(z=0,t) - T_{\text{atm}})
\end{equation}
\end{linenomath*}

\noindent where $Q_{0}$ is the amplitude of the periodic flux, and $a$~(W$\cdot$m$^{-2}$$\cdot$K$^{-1}$) is the atmospheric damping coefficient. Over land surfaces, as assumed in dry GCM configurations, an additional lower boundary condition must be applied to the subsurface; for example:

\begin{linenomath*}
\begin{equation}
T(z \gg \delta)= \overline{T}
\end{equation}
\end{linenomath*}

\noindent where $\overline{T}$ is some mean temperature in the deep subsurface, and $\delta=\sqrt{\frac{2k}{\omega \rho c_p}}$ (m) is the thermal skin depth corresponding to a forcing frequency of $\omega$ (s$^{-1}$), whether it be diurnal or annual. The magnitude of surface temperature fluctuations is expressed as (see Appendix~A for the full derivation):

\begin{linenomath*}
\begin{equation}
\Delta T(z=0) \approx \frac{2Q_0}{\sqrt{\rho c_p k} \sqrt{\omega} + a_{0}} \approx \frac{\Delta Q}{\sqrt{\rho c_p k} \sqrt{\omega} + a_{0}}
\end{equation}
\end{linenomath*}

\noindent where $\Delta Q$ denotes the amplitude of fluctuations in $Q_\text{net}$. The thermal inertia $I = \sqrt{\rho c_p k}$, combined with $\sqrt{\omega}$ represents damping of the energy (W$\cdot$m$^{-2}$$\cdot$K$^{-1}$) received at the surface, due to subsurface heat conduction. Meanwhile, $a_0$, which is the magnitude of $a$, accounts for damping due to direct energy exchange between the surface and the atmosphere.

For Titan, the temperature of the overlying atmosphere, $T_{\text{atm}}$, can be reasonably assumed to remain constant due to its long radiative timescale \citep{Flasar81, Flasar14}. Notably, the linear atmospheric damping term in Eq.~(4) takes the same form as the expression for sensible heat flux:

\begin{linenomath*}
\begin{equation}
a_{0} \approx \rho_a C_a C_H |V| \approx 5 \cdot 1000 \cdot 0.001 \cdot \mathcal{O}(10^{-2} \text{--} 10^{-1}) \approx \mathcal{O}(10^{-2} \text{--} 10^{-1})
\end{equation}
\end{linenomath*}

\noindent where $\rho_a$ (kg$\cdot$m$^{-3}$) is the density of Titan's lower atmosphere as measured by the Huygens probe, $C_a$ is the specific heat capacity of the N$_2$-dominated atmosphere (J$\cdot$kg$^{-1}\cdot$K$^{-1}$), $C_H$ is the surface heat transfer coefficient used in GCMs, and $|V|$ (m $\cdot$ s$^{-1}$) is the typical magnitude of the near-surface horizontal winds in GCM simulations 
\citep[e.g.,][]{Lora19}. In the following sections, we examine how the surface temperature variations reproduced in our GCM simulations at diurnal and seasonal timescales correspond to the theoretical predictions, Eq~(6).


\subsection{Titan Atmospheric Model}\label{Model}

We investigate the diurnal and seasonal variations of surface temperature and PBL height using the Titan Atmospheric Model (TAM), a three-dimensional GCM of Titan that has been extensively validated against observational data \citep{Lora15,Lora24}. TAM demonstrates particularly strong performance in reproducing lower and middle atmospheric temperatures when compared to observations \citep{Schinder11, Schinder12, Schinder20, Jennings16, Jennings19}. The model optionally incorporates one of two PBL parameterization schemes: a non-local K-profile scheme \citep{Holtslag93} or the Mellor–Yamad scheme \citep{Mellor82}. In this study, we employ the latter scheme, consistent with the approach taken by \cite{Charnay12}.

While TAM has demonstrated robust performance in simulating Titan's climate system, including the methane cycle \citep{Lora15, LA17, Faulk20, Battalio22, Lora22, Lora24}, we employ a simplified configuration with dry, flat, and homogeneous surface conditions to facilitate direct comparisons with previous studies and isolate the effects of thermal inertia on surface temperature variations. This simplified approach results in surface temperatures that exceed observed brightness temperatures due to the absence of evaporative cooling \citep{Mackenzie19b}. Consequently, our results should be interpreted as an investigation of temperature and PBL variation mechanisms rather than a representation of conditions on Titan as observed by Cassini.

Our investigation encompasses four simulation cases (Table~\ref{tab:table1}). Cases 1, 2, and 3 employ a uniform volumetric heat capacity of $\rho c_p = 3 \times 10^6$ J$\cdot$m$^{-3}\cdot$K$^{-1}$, with varying thermal conductivities of $k = 0.01$, $0.1$, and $1.0$ W$\cdot$m$^{-1}\cdot$K$^{-1}$, respectively, corresponding to thermal inertia values of $I = 173$, $548$, and $1,732$ TIU. Case~4 adopts a spatially varying thermal inertia map based on the ``convective lakes" scenario of \citet{Mackenzie19b}, which characterizes Titan’s surface using Cassini RADAR observations and classifies it into five terrain types: dunes, hummocky, labyrinth, plains, and lakes, which are assumed to be convecting. In our study, we fix $\rho c_p$ across Cases 1–3 as the volumetric heat capacity values across terrain types in \cite{Mackenzie19b} are relatively similar in magnitude. In contrast, thermal conductivity spans over two orders of magnitude, and our three selected values are chosen to broadly represent this range.


The simulations are run at T21 horizontal resolution (64 × 32 longitude-latitude grid) with 50 atmospheric vertical levels and 15 subsurface vertical levels, using a 600~s time step. A shortwave albedo of 0.2 and surface thermal emissivity of 0.95 are adopted. The radiative transfer parameterization in TAM, which uses plane-parallel two stream approximations, is non-gray and accounts for multiple scattering, employing absorption coefficients and haze optical properties from observations \citep{Lora15} and producing radiative heating rates that closely match estimates based on Huygens measurements \citep{Tomasko08c}. Still, the shortwave heating at high zenith angles (dawn and dusk) should be regarded with caution given the plane parallel assumption, though this issue is reduced in the thick and strongly scattering lower atmosphere. Near the equator, where the Huygens and Dragonfly landing sites are located, the horizontal resolution is about 253~km in both latitude and longitude, corresponding to an area of approximately $6.4 \times 10^{4}$ km$^{2}$ per grid cell. The atmospheric vertical resolution in the lower few kilometers is maintained at several tens of meters to sufficiently resolve boundary layer variations. Subsurface layer depths follow $z(j=1 \ \text{to} \ 10)=\delta_{d}/16 \cdot 2^{j-1}$ and $z(j=11 \ \text{to} \ 15)=z(j-1)+16\delta_{d}$ , where $\delta_{d}$ represents the diurnal skin depth for Titan and $j$ denotes the subsurface level. This configuration ensures adequate resolution for both diurnal and seasonal subsurface heat conduction by progressively increasing layer depth, with 5 layers within $\delta_d$, 10 layers within $\delta_a$ (the annual skin depth), and 5 layers extending to the deepest region at $z(j=15) \approx 4.3\delta_a$. 

The model is initialized from a cold start with uniform surface and subsurface temperatures of $T = 94$~K, and integrated for 200 Titan years with diurnally averaged output. For diurnal analyses, the simulation is extended by one additional Titan year with output saved every 16 Earth hours (i.e., 1/24 of a Titan day), focusing on a 30 Titan-day period centered around southern midsummer in the final year.


\section{Surface Temperature}
\subsection{Diurnal Variations}

Figure~\ref{fig:diurnal_1d_huygens_EB} presents the diurnal evolution of surface energy balance components and surface temperature near the Huygens landing site during midsummer for the four simulation cases. In Case~4, the plains terrain type ($I=601$~TIU) is assumed at the Huygens landing site \citep{Mackenzie19b}. In Case~1, the net shortwave (SW) flux increases from 0.0 to approximately 1.3~W$\cdot$m$^{-2}$ from nighttime to local noon, while the sensible heat flux peaks near 1.2~W$\cdot$m$^{-2}$ around midday, indicating that a substantial portion of incoming energy is returned to the atmosphere. This behavior reflects the limited capacity of low thermal inertia surfaces to conduct heat into the subsurface, necessitating that excess surface energy be dissipated through sensible heat flux. During the evening and nighttime, the sensible heat flux turns slightly negative, indicating rapid surface cooling and the formation of a near-surface temperature inversion. The inversion is evident in the potential-temperature structure and is discussed in Section~4.1. As thermal inertia increases (Cases 2 and 3), the daytime sensible heat flux amplitude decreases to 1.0 and 0.8~W$\cdot$m$^{-2}$, as subsurface heat conduction more effectively removes energy from the surface. During the evening and nighttime, the sensible heat flux shifts toward positive values, reflecting slower surface cooling. Meanwhile, the net infrared (IR) radiative flux remains relatively constant at approximately 0.07~W$\cdot$m$^{-2}$ across all cases, consistent with the relatively small changes in surface temperature. Case~4 shows similar behavior to Case~2 due to their comparable thermal inertias.

The simulated diurnal surface temperature variations are approximately $\Delta T_d(\text{TAM}) \approx 0.21 \pm 0.02$, $0.18 \pm 0.02$, and $0.12 \pm 0.02$~K for Cases~1, 2, and 3, respectively. These values represent the mean and one standard deviation computed over a 30 Titan-day period centered on midsummer ($L_s = 290^\circ$–$305^\circ$) in the final simulation year. For a diurnal forcing frequency of $\omega_{d} = \frac{2\pi}{16 \times 86400}$~s$^{-1}$ and net energy flux amplitudes of $\Delta Q_d(\text{TAM}) \approx 0.08 \pm 0.01$, $0.19 \pm 0.02$, and $0.48 \pm 0.02$~W$\cdot$m$^{-2}$ for Cases~1, 2, and 3, respectively—derived using the same 30 Titan-day averaging method as for the simulated surface temperature amplitudes—and with an atmospheric damping coefficient of $a_0 \approx 0.10$~[W$\cdot$m$^{-2}\cdot$K$^{-1}$], obtained as the 30-day mean value, the theoretical temperature variations according to Eqs.~(6) and (7) are estimated to be $\Delta T_d(\text{theory})=0.20$, 0.16, and 0.13~K. These estimates closely align with the simulated values, reinforcing the validity of the theoretical framework despite its simplification of the net energy flux as a sine-like function. For Case 4, the $\Delta Q_d(\text{TAM}) \approx 0.24 \pm 0.02$~W$\cdot$m$^{-2}$ yields a theoretical expectation of $\Delta T_d(\text{theory}) \approx 0.18$~K, closely matching the $\Delta T_d(\text{TAM}) \approx 0.18 \pm 0.02$~K as well.

Despite the fact that the net energy flux—the residual of the surface energy budget—increases with thermal inertia, the amplitude of surface temperature variations decreases. The decreasing amplitude of temperature variation arises because increasing thermal inertia enhances subsurface heat conduction, effectively buffering the surface temperature despite the higher available energy. The values of $I \cdot \sqrt{\omega_d}$ corresponding to the thermal inertias used in Cases~1, 2, 3 and 4 are approximately 0.37, 1.17, 3.69 and 1.28~W$\cdot$m$^{-2}\cdot$K$^{-1}$, respectively. These values significantly exceed the atmospheric damping coefficient, $a_0 \approx 0.10$~W$\cdot$m$^{-2}\cdot$K$^{-1}$, indicating that subsurface heat conduction is the dominant process in regulating surface temperature variations at diurnal timescales.

\subsection{Seasonal Variations}

Figure~\ref{fig:seasonal_temp} shows the annual zonal-mean surface temperatures as a function of latitude and time for the four simulation cases. As previously reported in \citet{Mackenzie19b}—which explored cases using a uniform thermal inertia of $I = 335$ TIU \citep[representing porous ice regolith;][]{Tokano05}, a globally averaged value of $I = 750$ TIU, and a spatially varying thermal inertia based on the ``convective lakes" scenario—seasonal surface temperatures in our Cases 1, 2, and 4 similarly exhibit little to no sensitivity to thermal inertia. In Case~4, the surface is dominated by low thermal inertia terrains: Plains make up approximately 65\% of the surface area ($I = 601$ TIU), dunes 18\% ($I = 236$ TIU), and labyrinth terrain 3\% ($I = 808$ TIU). In contrast, high thermal inertia terrain—hummocky regions with $I = 1{,}962$ TIU—comprise only 14\% of the surface \citep[see Figure~2 from ][]{Mackenzie19b}. Among our simulation cases, Case~3 alone exhibits a distinct structure compared to the other three, indicating that seasonal surface temperature variations begin to show a noticeable sensitivity to thermal inertia only when the thermal inertia is sufficiently high globally. 

This behavior can be qualitatively understood using Eqs.~(6) and (7). For an annual forcing frequency of $\omega_{a} = \frac{2\pi}{29.5 \times 365 \times 86400}$~s$^{-1}$, the corresponding values of $I \cdot \sqrt{\omega_a}$ for Cases~1, 2, 3, and 4 are approximately 0.014, 0.045, 0.14, and 0.049~W$\cdot$m$^{-2}\cdot$K$^{-1}$, respectively. Among these, only Case~3 exceeds the atmospheric damping coefficient $a_0 \approx 0.05$ W$\cdot$m$^{-2}\cdot$K$^{-1}$, obtained as the annual mean, indicating that subsurface heat conduction becomes significant only when thermal inertia is sufficiently high. On seasonal timescales, subsurface conduction is generally less important than atmospheric damping primarily due to Titan’s long orbital period (29.5 Earth years), which yields a small $\omega_a$. This stands in contrast to the diurnal regime, wherein subsurface heat conduction can dominate the surface temperature variability at intermediate values of thermal inertia.

Figure~\ref{fig:seasonal_1d_huygens_EB} presents the seasonal evolution of surface energy balance components and surface temperature near the Huygens landing site for the four simulation cases. Across all cases, the net SW flux varies seasonally from approximately 0.20 to 0.37~W$\cdot$m$^{-2}$, increasing from winter through summer. The sensible heat flux exhibits a broadly similar seasonal pattern, ranging from 0.15 to 0.30~W$\cdot$m$^{-2}$. Lower-thermal-inertia cases exhibit larger fluctuation amplitudes, driven by greater surface temperature variability. The net IR flux remains relatively steady at approximately 0.07~W$\cdot$m$^{-2}$ across seasons and simulation cases. The resulting net energy flux is smaller in the lower thermal inertia cases, where a stronger sensible heat flux efficiently removes much of the incoming surface energy.

Surface temperature variations across seasons remain modest—under 1~K—and exhibit broadly similar patterns across all simulation cases. However, the amplitude of these variations is more strongly damped for higher thermal inertia. Applying Eqs.~(6) and (7) to the simulated seasonal net energy flux amplitudes of $\Delta Q_a(\text{TAM}) \approx 0.04 \pm 0.01$, $0.06 \pm 0.01$, and $0.10 \pm 0.02$~W$\cdot$m$^{-2}$ for Cases~1, 2, and 3, respectively, yields theoretical temperature variations of $\Delta T_a(\text{theory}) \approx 0.63$, 0.57, and 0.51~K. Here, $\Delta Q_a(\text{TAM})$ is derived from means calculated over the final 30 simulation years where 10-Titan-day moving averages are used to remove short-term variations. These estimates are roughly consistent with the simulated surface temperature variations of $\Delta T_a(\text{TAM}) \approx 0.95 \pm 0.04$, $0.80 \pm 0.04$, and $0.59 \pm 0.03$~K, though some discrepancies remain. These differences likely stem from the assumption of a constant $a_0$ across seasons, whereas in reality the horizontal wind speed $|V|$ exhibits seasonal variability in the simulations. Despite discrepancies between $\Delta T_a(\text{TAM})$ and $\Delta T_a(\text{theory})$ due to the simplification, Eqs.~(6) and (7) effectively capture the dominant role of atmospheric damping in regulating surface temperature variations at seasonal timescales. For instance, omitting the atmospheric damping term $a_0$ leads to theoretical predictions of $\Delta T_a(\text{theory}) \approx 2.81$, 1.22, and 0.70~K, which significantly overestimate the temperature response compared to the simulations. For Case~4, the $\Delta Q_a(\text{TAM}) \approx 0.06 \pm 0.01$~W$\cdot$m$^{-2}$ yields a theoretical surface temperature variation of $\Delta T_a(\text{theory}) \approx 0.57$~K, nearly identical to that of Case~2.

\section{Planetary Boundary Layer}
\subsection{Diurnal Variations}

\begin{figure}[htb!]
\centering
\includegraphics[width=0.82\textwidth]{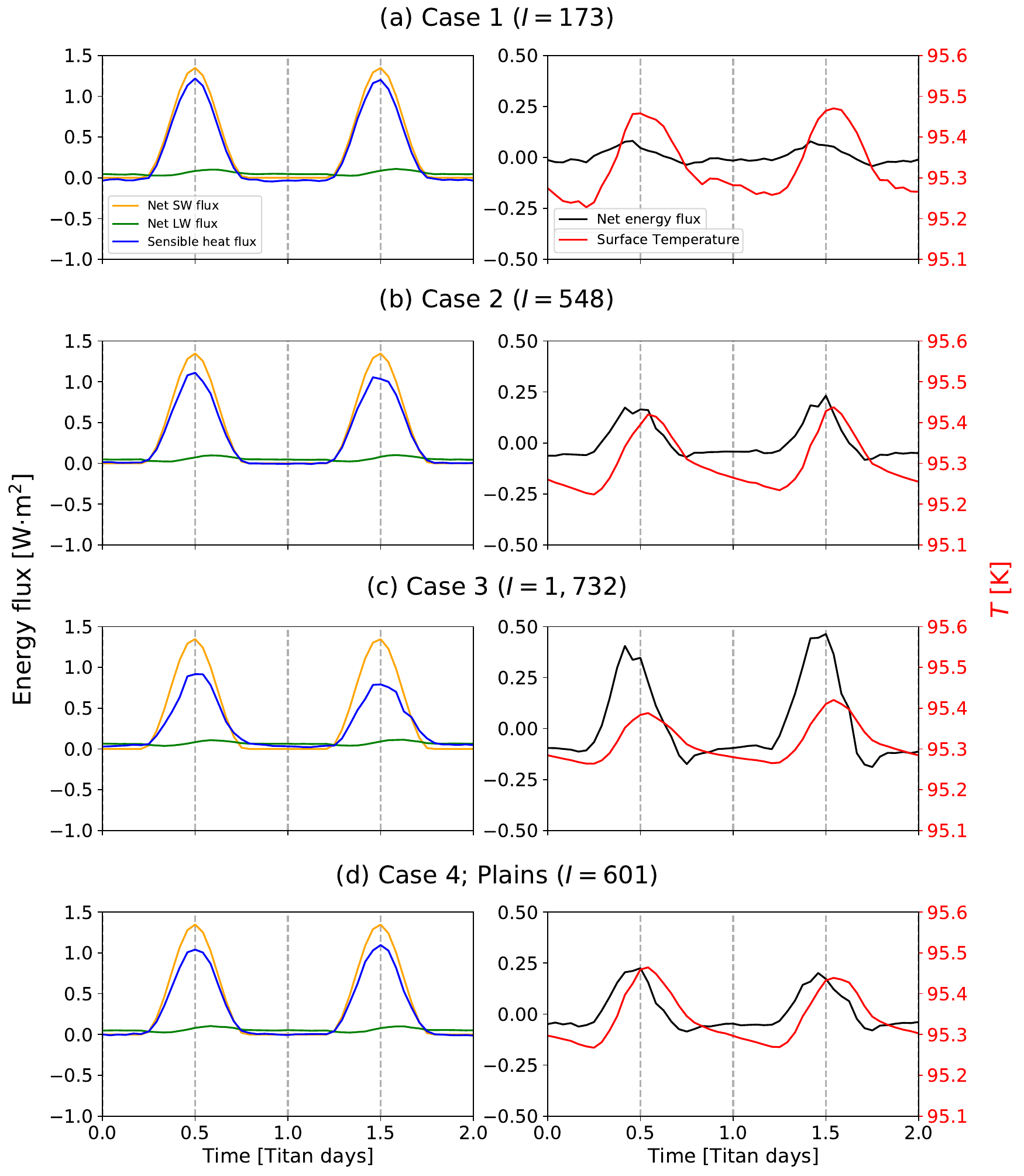}
\caption{Surface energy balance components and temperature variations at diurnal timescales at the model grid point nearest the Huygens landing site (167.7°E, 10.3°S) shown for four simulation cases with varying surface thermal inertia values. Panel (a): Case 1 ($I = 173$~TIU); (b) Case 2 ($I = 548$~TIU); (c) Case 3 ($I = 1,732$~TIU); and (d) Case 4 ($I=601$~TIU where plains terrain type is assumed \citep{Mackenzie19b}). TIU denotes Thermal Inertia Unit J$\cdot$m$^{-2}\cdot$s$^{-0.5}\cdot$K$^{-1}$. Results are sampled at 16-Earth-hour intervals (i.e., 1/24 of a Titan day) during midsummer conditions ($L_s = 300^\circ$, where $L_s$ is solar longitude) in the final simulation year. The time origin ($t = 0$) corresponds to local midnight. Net shortwave (SW) flux and net energy flux are defined as positive downward (into the surface), while net longwave (LW) flux and sensible heat flux are defined as positive upward (away from the surface). The simulated surface temperatures are approximately 2~K higher than the Huygens observations, primarily because our dry configuration lacks evaporative cooling. This warm bias persists throughout the simulations and is evident in other figures as well.}
\label{fig:diurnal_1d_huygens_EB}
\end{figure}

Figure~\ref{fig:diurnal_1d_huygens_PBL} shows the diurnal evolution of PBL height across the four simulation cases near the Huygens landing site. In each case, the left panels present snapshots of potential temperature profiles at 9:00~AM, 12:00~PM, and 3:00~PM local solar time (blue), as well as 12:00 AM (purple), compared with in-situ Huygens probe measurements \citep{Fulchignoni05} taken around 10:00~AM (black). Surface heating and the deepening of the adiabatic layer due to dry convection during the day are evident in all cases. At 12:00 AM, a stable near-surface inversion forms in Case 1, consistent with the negative sensible-heat flux in Figure~\ref{fig:diurnal_1d_huygens_EB}. As thermal inertia increases across cases, the 12:00 AM profiles become progressively less stable, consistent with reduced nocturnal cooling and a warmer surface. Notably, in Case~3 the 12:00 AM profile is similar to those at other local times, indicating that a near-surface adiabatic layer persists throughout the diurnal cycle under high thermal inertia. The middle panels show the corresponding temperature diffusion tendencies predicted by the model’s PBL parameterization, with arrows marking the profiles at 9:00~AM, 12:00~PM, and 3:00~PM corresponding to the inflection points in the left panels. The diffusion tendencies are generally positive, reflecting upward heat transfer by convection, but contain interspersed negative layers that create a noisy appearance. The evolution of these tendencies closely tracks the changes in the potential temperature profiles.

The right panels display the mean composite evolution of PBL depth, averaged over 30 Titan days during midsummer ($L_s = 290^\circ$–$305^\circ$). We diagnose PBL height as the height above ground where the vertical eddy diffusivity first falls below $0.01~\mathrm{m^2\,s^{-1}}$. This diffusivity-threshold criterion is the default in TAM when the Mellor–Yamada scheme is used and it closely delineates the top of the near-surface adiabatic layer. The time series is smoothed with a Savitzky–Golay filter (window = 5 bins, polynomial order = 2) to suppress bin-to-bin noise while preserving diurnal phase and amplitude. Results are insensitive to reasonable choices of window (3–7 bins).  
These diagnosed PBL heights reflect variations in surface heating and the resulting convective mixing, as corroborated by the inflection point of the potential temperature profile and the temperature diffusion tendencies in the left and middle panels. Although the individual PBL-depth time series show variability with occasional outliers (e.g., Case~2), these are attributable to natural variability; averaging effectively suppresses them.

For the lowest thermal inertia surface (Case~1), the PBL grows from $130\pm90$~m to $1{,}130\pm400$~m by mid-afternoon, where uncertainty ranges correspond to one standard deviation from the mean. With increasing thermal inertia, baseline nighttime PBL depths rise to $270\pm130$~m and $600\pm190$~m, and maximum daytime depths reach $1{,}220\pm370$~m and $1{,}420\pm290$~m for Case~2 and Case~3, respectively. The significant increase in nighttime PBL depths for higher thermal inertia cases reflects reduced nighttime surface cooling, which allows the PBL to remain deeper (Figure~\ref{fig:diurnal_1d_huygens_EB}). For Case~4, the PBL grows from $290\pm130$~m to $1{,}290\pm330$~m by mid-afternoon, closely matching Case~2, which employs a comparable thermal inertia. Daytime maximum PBL depths are deeper for higher-thermal-inertia surfaces, despite the smaller daytime sensible heat fluxes over these surfaces. This likely reflects that daytime depth is controlled by the integrated heat accumulation over the day rather than by instantaneous fluxes. Indeed, the PBL fluctuation amplitude—reflecting the magnitude of PBL growth—is greatest for the lowest thermal inertia case, consistent with its larger sensible heat flux variability. Differences in sensible heat flux lead to noticeable variations near the tropopause ($\sim$100~mbar, $\sim$50~km altitude): Case~1 shows a higher temperature ($\sim$71~K) than Case~3 ($\sim$70~K), with Case~2 and 4 intermediate (not shown). This reflects the cumulative effect of sensible heat flux over time: low thermal inertia surfaces transfer more energy upward, warming the free atmosphere and raising tropopause temperature, whereas high thermal inertia surfaces limit this warming.

\subsection{Seasonal Variations}

\begin{figure}[htb!]
\centering
\includegraphics[width=0.78\textwidth]{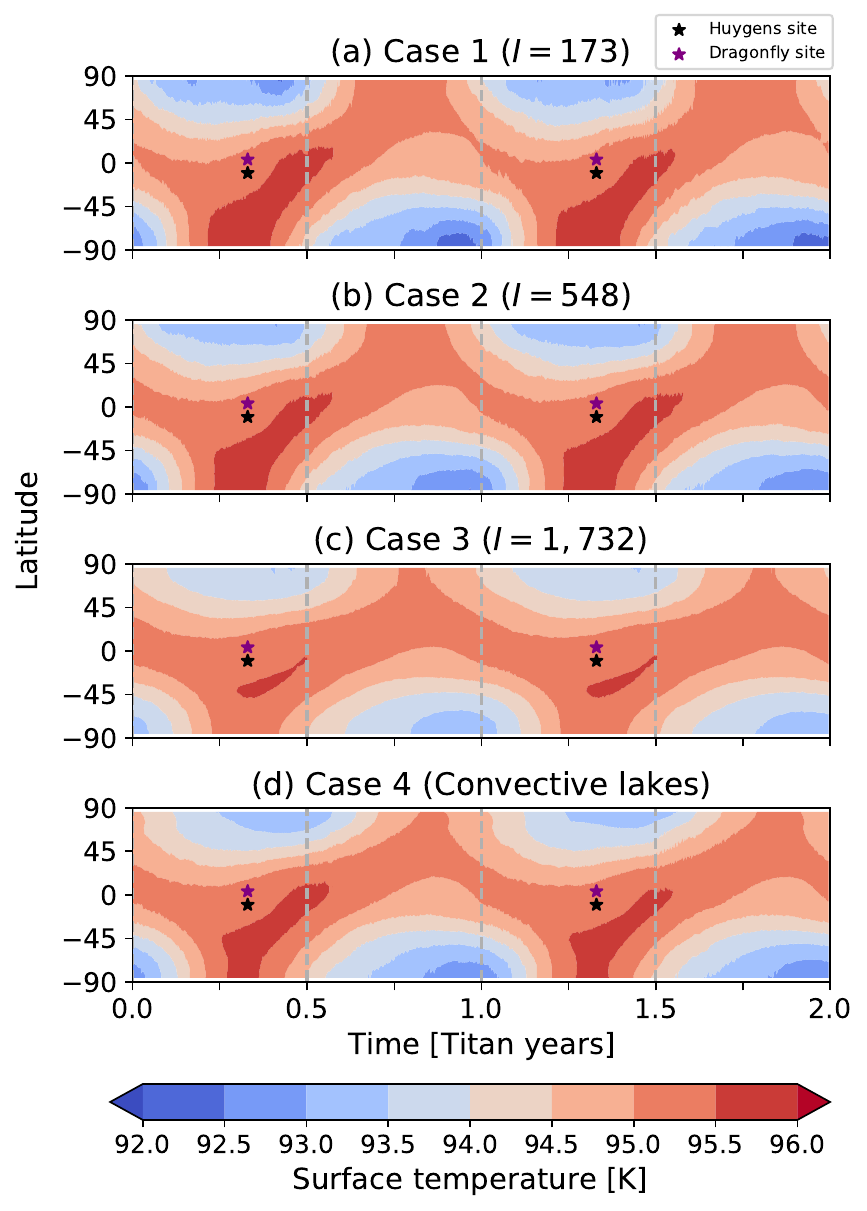}
\caption{Simulated zonal-mean surface temperature as a function of latitude and time for four simulation cases. Cases~1–3 assume globally uniform thermal inertia, while Case 4 adopts a spatially varying thermal inertia map based on the convective lakes scenario from \citet{Mackenzie19b}. Time is referenced to southern spring equinox ($t = 0$ corresponds to $L_s = 180^\circ$). Each panel displays the final two years of a 200-year simulation, reflecting equilibrated seasonal temperature patterns. The corresponding latitudes and seasonal timings of the Huygens and Dragonfly landing sites are marked with stars.}
\label{fig:seasonal_temp}
\end{figure}

\begin{figure}[htb!]
\centering
\includegraphics[width=0.85\textwidth]{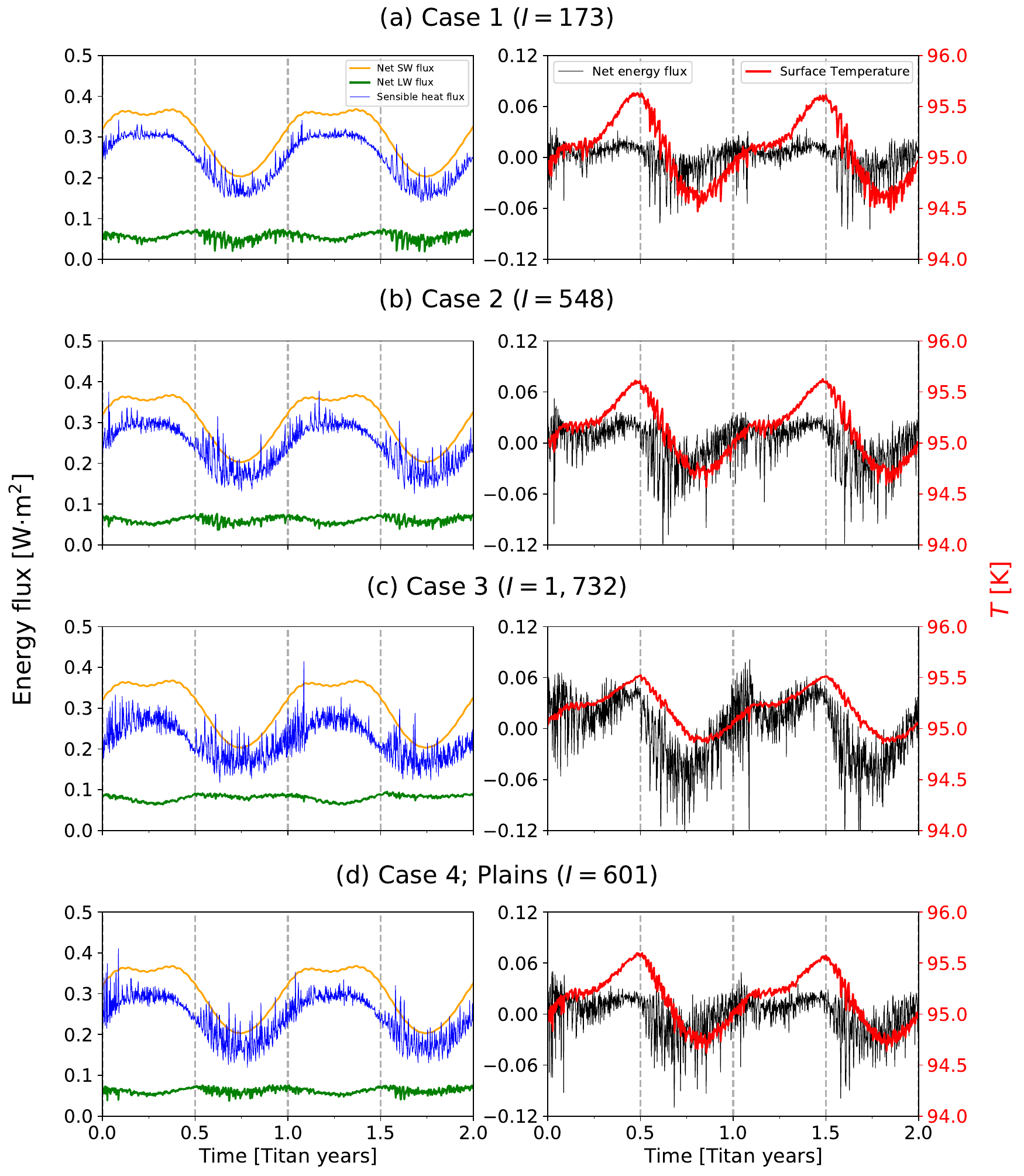}
\caption{Similar to Figure~\ref{fig:diurnal_1d_huygens_EB}, but for seasonal timescales, where diurnally averaged values are plotted. Results are taken from the final two years of the 200-year simulations, representing equilibrated seasonal cycles at the Huygens landing site. The seasonal phases corresponding to the Huygens and Dragonfly landing sites occur at $t=0.33$.}
\label{fig:seasonal_1d_huygens_EB}
\end{figure}

\begin{figure}[htb!]
\centering
\includegraphics[width=1.0\textwidth]{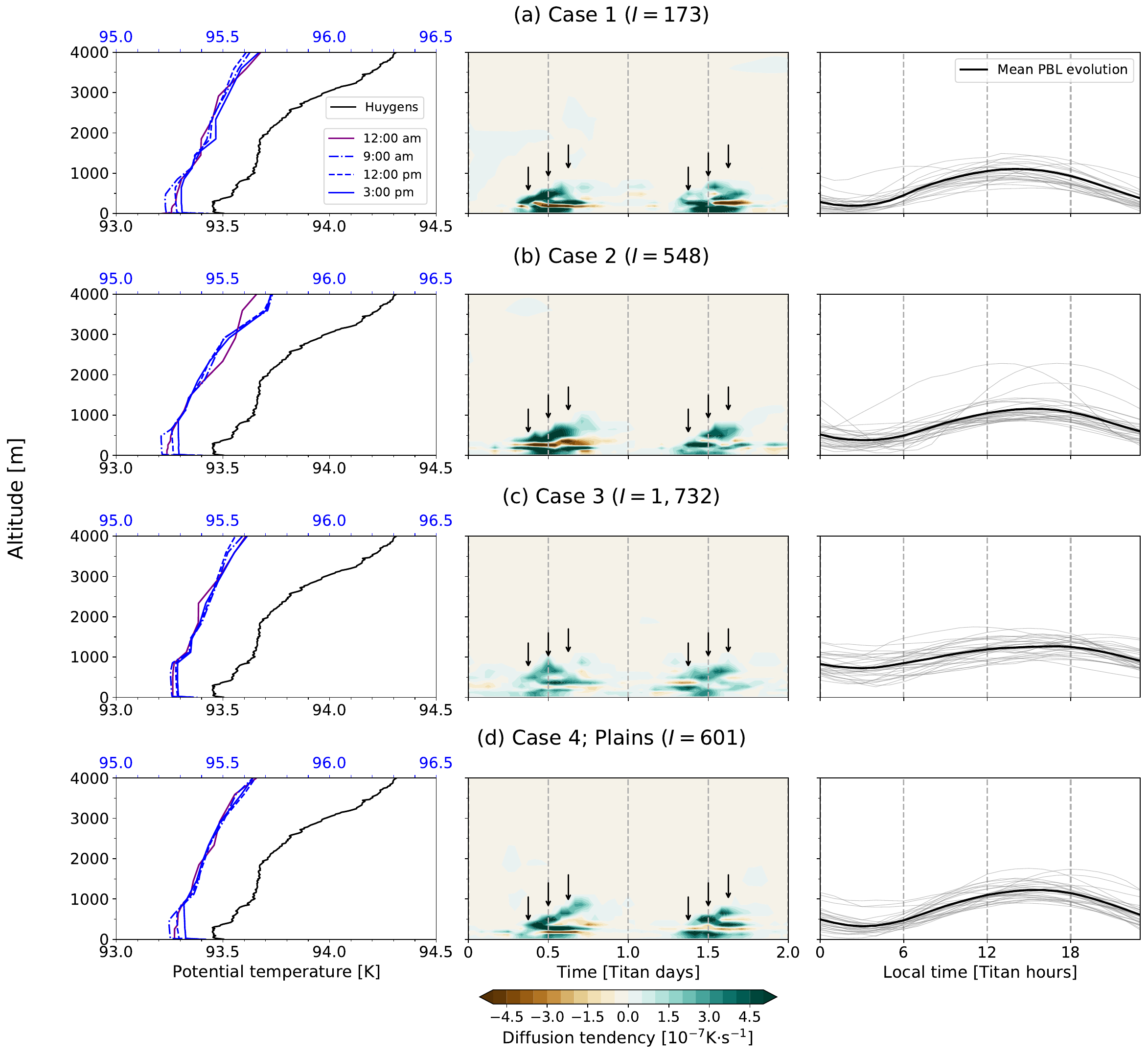}
\caption{Diurnal variations in boundary layer depth for the four simulation cases. The left panels for each case show the diurnal evolution of potential-temperature profiles at 9:00 AM, 12:00 PM, and 3:00 PM local solar time (blue) and at 12:00 AM (purple), compared with the Huygens profile obtained at approximately 10:00 AM \citep{Fulchignoni05}. The middle panels display the temperature diffusion tendency predicted by the model’s PBL parameterization, with arrows indicating the profiles at 9:00 AM, 12:00 PM, and 3:00 PM, corresponding to the inflection points in the left panels. The right panels show the mean composite evolution of PBL depth, averaged over 30 Titan days during midsummer ($L_s = 290^\circ$–$305^\circ$), with smoothing applied to highlight the diurnal amplitude. The mean curve is shown as a thick line, and individual days are shown as thin lines. The PBL height is diagnosed as the altitude of the model layer where the vertical diffusion coefficient first drops below a threshold of 0.01~m$^2$$\cdot$s$^{-1}$, which approximately corresponds to the inflection point of the potential temperature profile in the boundary layer parameterization in TAM.}
\label{fig:diurnal_1d_huygens_PBL}
\end{figure}

\begin{figure}[htb!]
\centering
\includegraphics[width=0.95\textwidth]{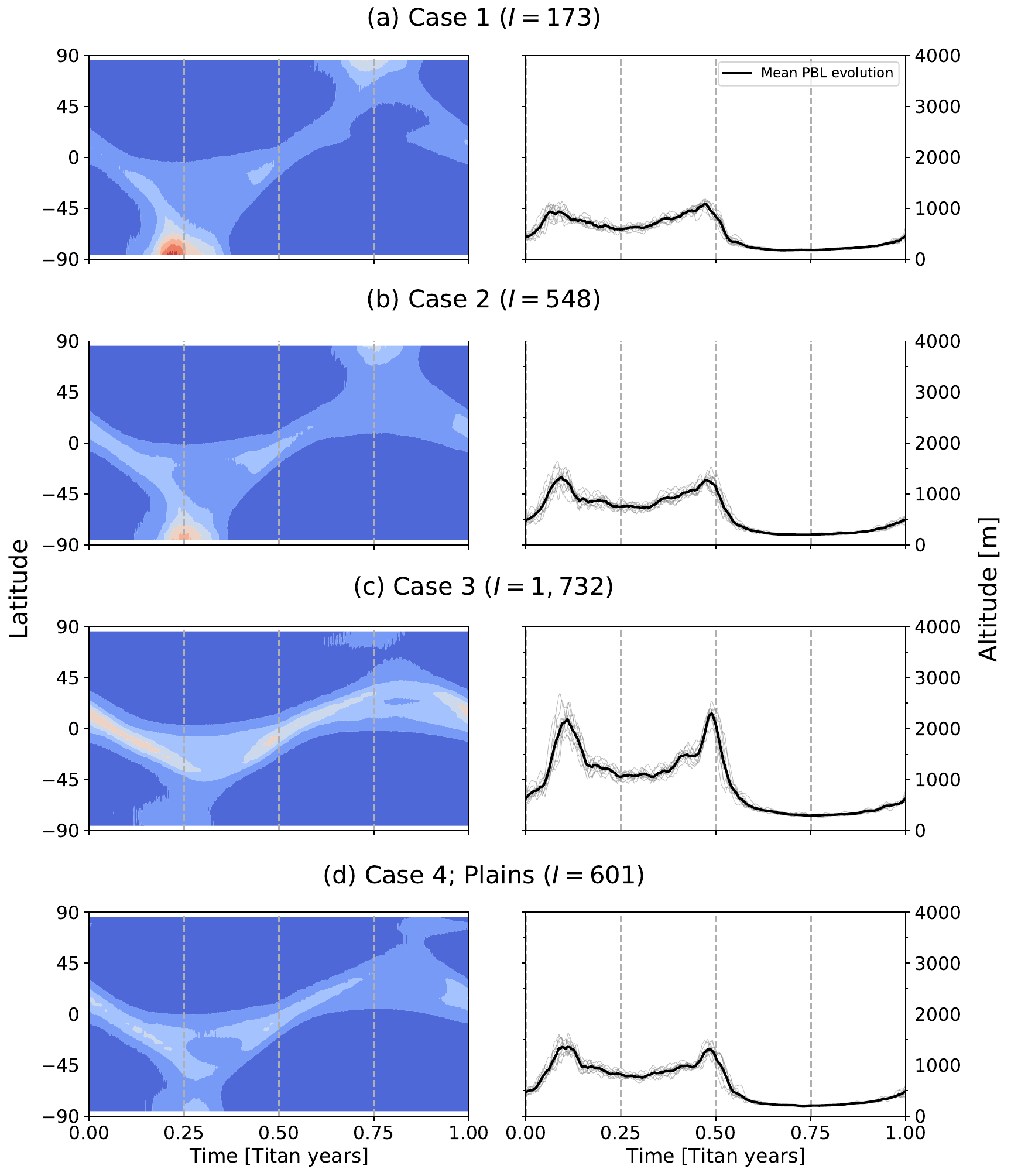}
\caption{Seasonal variations of PBL depth for four simulation cases. The left panels show global zonal-mean fields, averaged over 10 Titan years; the right panels show the composite evolution of PBL depth, also averaged over 10 Titan years and smoothed to emphasize the seasonal amplitude. The mean curve is shown as a thick line, and individual years are shown as thin lines.}
\label{fig:seasonal_PBL}
\end{figure}

Figure~\ref{fig:seasonal_PBL} presents the seasonal evolution of the PBL height across the four simulation cases. The left panels in each case display global zonal means compiled over 10 Titan years, while the right panels show the 10-Titan-year mean composite PBL depth near the Huygens landing site, where Savitzky–Golay smoothing (30 Titan-day window, third order) is applied to emphasize the seasonal amplitude. The global seasonal structure of the PBL closely mirrors the surface temperature patterns shown in Figure~\ref{fig:seasonal_temp}, indicating that PBL depth variations are primarily governed by the seasonal migration of regions with peak surface heating. The hemispheric asymmetry in the PBL structure is evident, as it is in the surface temperature, due to orbital eccentricity. 

In Case~1, the diagnosed PBL height at the Huygens landing site ranges from approximately 180~m during southern winter to about 1,100~m between the southern vernal ($t \approx 0.0$) and autumnal ($t \approx 0.5$) equinoxes. In Case~2, it ranges from 200~m to 1,320~m, while in Case~3, it extends from 290~m to 2,300~m. The increase in baseline (minimum) PBL depths for higher thermal inertia cases is modest, whereas the increase in maximum depths is more pronounced. In both cases, the elevated PBL heights for higher thermal inertia simulations are consistent with the diurnal-timescale behavior: a reduced contrast between minimum and maximum surface temperatures produces a more steady temperature structure that allows greater energy buildup and formation of a deeper PBL.

\begin{figure}[htb!]
\centering
\includegraphics[width=1.0\textwidth]{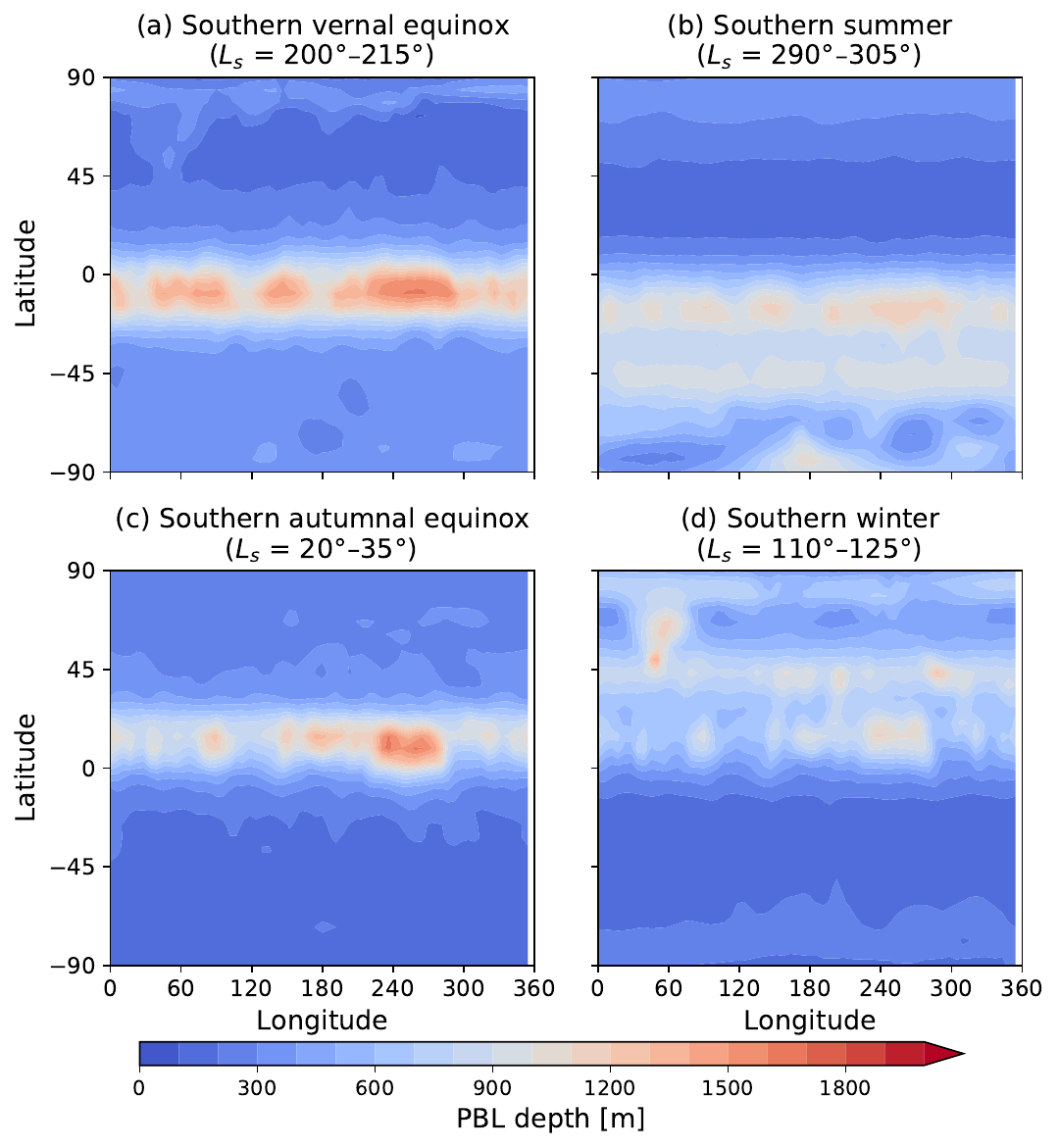}
\caption{
Global seasonal mean maps of diagnosed PBL depth for four distinct seasons, based on simulations with a spatially varying thermal inertia map (Case~4). (a) Southern spring, averaged over $L_s=200^\circ$–$215^\circ$; (b) Southern summer, $L_s=290^\circ$–$305^\circ$; (c) Southern autumn, $L_s=20^\circ$–$35^\circ$; and (d) Southern winter, $L_s=110^\circ$–$125^\circ$.
}
\label{fig:seasonal_PBL_map}
\end{figure}

Figure~\ref{fig:seasonal_PBL_map} presents global seasonal mean maps of the diagnosed PBL depth for four distinct seasons, using Case~4 to illustrate how local variations in thermal inertia influence PBL depth. The maximum depth is around 1,800~m, and structural differences across longitudes are apparent due to spatial variations in thermal inertia. For instance, an elevated PBL depth is observed in the Xanadu region (approximately 250°E near the equator) across the seasons, where the assumed thermal inertia corresponds to hummocky terrain (1,962~TIU), compared to dune fields (236~TIU) and plains (601~TIU) at similar latitudes elsewhere. The imprint of this enhancement is especially pronounced during the southern summer and winter solstices. Additionally, elevated PBL depths are evident in polar regions—near 90°S, 180°E during southern summer and near 70°N, 40°E during southern winter, attributable to the presence of ``convective lakes'' with extremely high thermal inertia (18,090~TIU) \citep{Mackenzie19b}. These results suggest that locally varying thermal inertia strongly influences the seasonal evolution of the PBL depth.

\begin{figure}[htb!]
\centering
\includegraphics[width=0.73\textwidth]{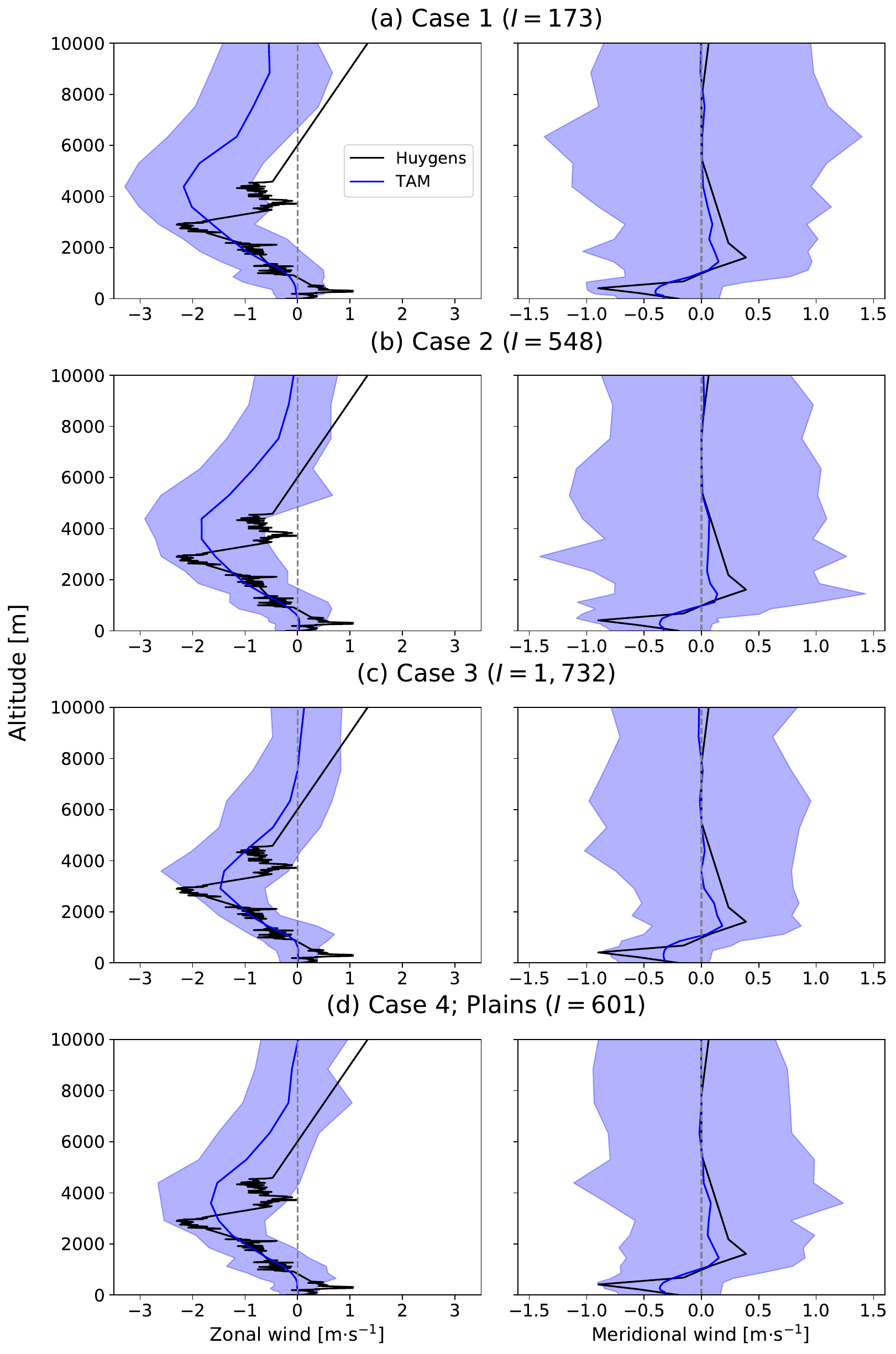}
\caption{Zonal and meridional wind profiles at the model grid point nearest the Huygens landing site for the four simulation cases compared with Huygens probe observations \citep{Bird05, Karkoschka16}. Positive values in the zonal and meridional directions correspond to eastward and northward motions, respectively. Solid blue lines indicate seasonal mean profiles, averaged over 30 Titan days during midsummer in the final simulation year. Shaded regions denote the 3rd to 97th percentile range of simulated winds.}
\label{fig:seasonal_1d_huygens_winds}
\end{figure}

Figure~\ref{fig:seasonal_1d_huygens_winds} shows the zonal and meridional wind profiles at the model grid point closest to the Huygens landing site for the four simulation cases, compared with the Huygens probe observations \citep{Bird05, Karkoschka16}. Positive values in the zonal and meridional directions correspond to eastward and northward motions, respectively. The seasonal mean is computed as a 30 Titan-day average centered around midsummer. During the period, the simulated wind profiles exhibit minimal diurnal variation (not shown). Across all cases, the simulations reproduce the observed wind structure in the lowest few kilometers, successfully capturing the key features of both the zonal and meridional components. The simulated near-surface wind magnitudes are on the order of $\mathcal{O}(10^{-2})$–$\mathcal{O}(10^{-1})$~m$\cdot$s$^{-1}$, consistent with results from previous Titan model intercomparison studies \citep{Lora19}. While the simulated near-surface winds are centered near zero on average, the model does occasionally reproduce eastward winds consistent with observations; however, the strongest observed eastward values lie outside the model variability. This residual bias may reflect, at least in part, the omission of convective vertical momentum transport, though it could also indicate that the Huygens observation was atypical. As shown in \citet{Charnay15}, equatorial tropical methane storms during equinox can transport eastward winds and their associated momentum from higher altitudes downward, potentially contributing to the eastward orientation of the dune fields \citep{Lorenz06, Lorenz09}.

Titan’s middle atmosphere exhibits eastward superrotation driven by upward and equatorward angular momentum transport, while friction keeps near-surface flow close to solid-body rotation; the resulting low-latitude momentum budget favors weak mean westward winds near the surface, which our simulations reproduce and which agree with prior Titan GCM studies \citep{Flasar09,Lewis23,Lombardo23b,Lora19}. Above $\sim$4 km in the troposphere, however, our simulated zonal winds are systematically weaker than observed. This bias likely reflects differences in the simulated thermal structure: by thermal-wind balance, understated zonal temperature gradients yield weaker shear. In our dry runs, the zonal surface-temperature gradients peak toward the poles (Fig.~\ref{fig:seasonal_temp}), whereas including moist processes leads to more realistic gradients and winds \citep{Lora19}. Although near-surface differences among Cases 1–4 are small, the stratospheric response varies substantially (not shown): peak speeds reach $\sim150~\mathrm{m\,s^{-1}}$ in Cases 1, 2, and 4 but only $\sim60~\mathrm{m\,s^{-1}}$ in Case 3, suggesting that global surface-temperature structure—set here by thermal inertia—can imprint on the circulation aloft. A full analysis of the vertical momentum pathways responsible for this sensitivity is left for future work, while our focus here is the lowest few kilometers where boundary-layer processes dominate and model–data agreement is generally good.

For meridional winds, the flow near the surface is predominantly southward, transitioning to northward at altitudes around 2~km. This vertical wind reversal suggests the presence of a cross-equatorial Hadley circulation, characterized by rising motion in the southern summer hemisphere and descending motion in the northern winter hemisphere. As a result, there is northward return flow aloft and a compensating southward return flow near the surface at the latitude of the Huygens landing site \citep{Lora25}. Wind magnitudes—both zonal and meridional—are generally smaller in the higher thermal inertia cases, although the differences are minimal. This likely reflects weaker temperature gradients, as higher thermal inertia surfaces dampen  surface temperature contrasts (Figure~\ref{fig:seasonal_temp}), thereby reducing the thermal forcing that drives the atmospheric circulation.

\section{PBL Characteristics at the Dragonfly Landing Site}

In this section, we present the same analysis of surface temperature and PBL structure as in the previous sections, but for the planned Dragonfly landing site near Selk Crater ($161.8^\circ$E, $3.7^\circ$N) during southern midsummer ($L_s \approx 300^\circ$) \citep{Barnes21, Lorenz21a}. For Case~4, the dunes terrain type, with a thermal inertia of $I=236$~TIU, is assumed at the surface \citep{Mackenzie19b}.

\begin{figure}[htb!]
\centering
\includegraphics[width=1.0\textwidth]{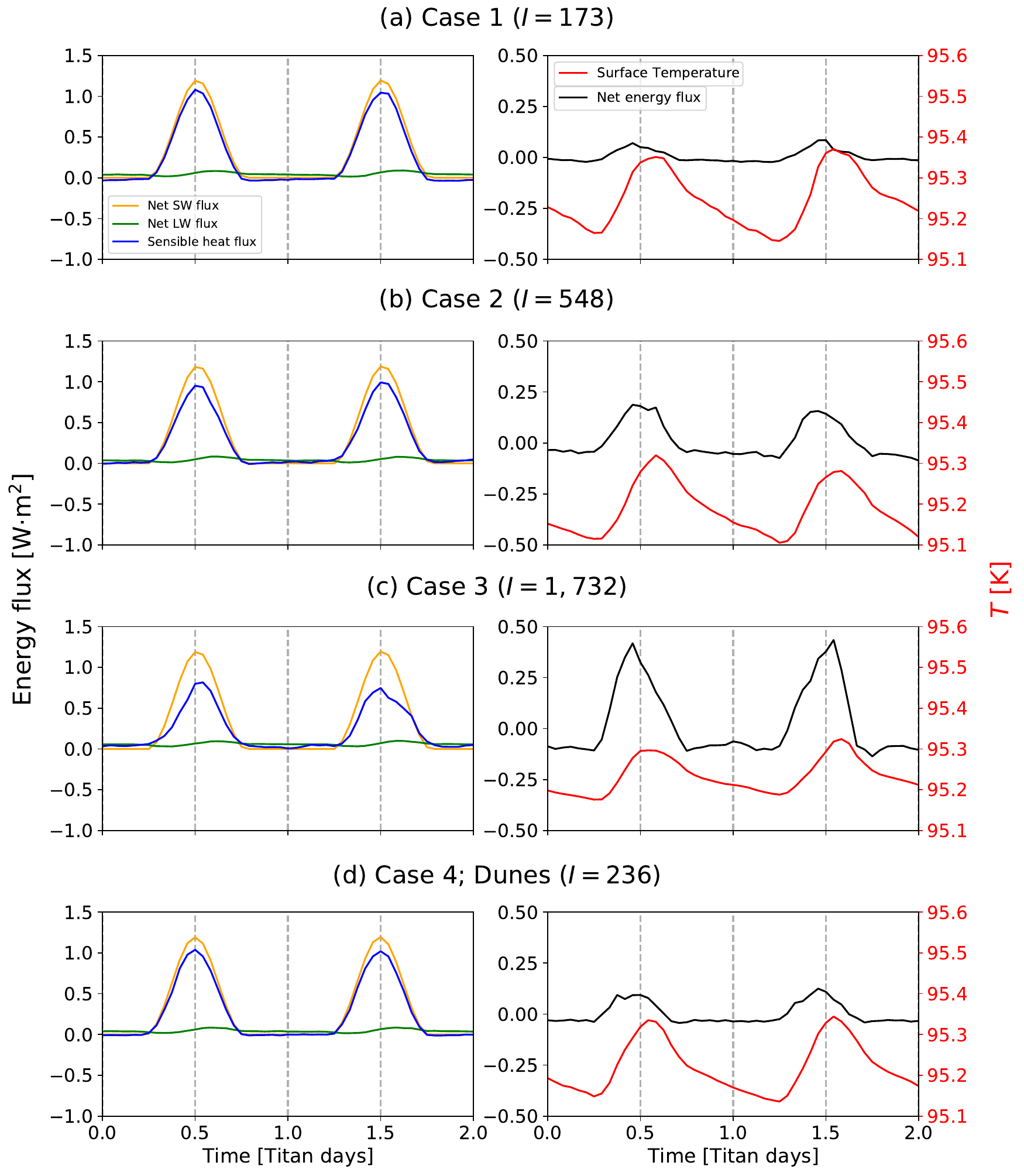}
\caption{Similar to Figure~\ref{fig:diurnal_1d_huygens_EB}, but for the model grid point nearest the planned Dragonfly landing site (161.8°E, 3.7°N), rather than the Huygens landing site. For Case 4, the dunes terrain type, with a thermal inertia of $I=236$~TIU, is assumed at the surface \citep{Mackenzie19b}.}
\label{fig:diurnal_1d_dragonfly_EB}
\end{figure}

Figure~\ref{fig:diurnal_1d_dragonfly_EB} shows the diurnal energy balance components at the Dragonfly landing site. Maximum insolation is reduced compared to the Huygens landing site, as the Dragonfly site experiences northern winter conditions while the Huygens site corresponds to southern summer. This results in a lower mean surface temperature at the Dragonfly site, reduced by approximately 0.1~K relative to the Huygens site. Nevertheless, the amplitude of diurnal surface temperature variations remains comparable to that at the Huygens site. For Case~4, the simulated diurnal amplitude is $0.19 \pm 0.02$~K.

\begin{figure}[htb!]
\centering
\includegraphics[width=1.0\textwidth]{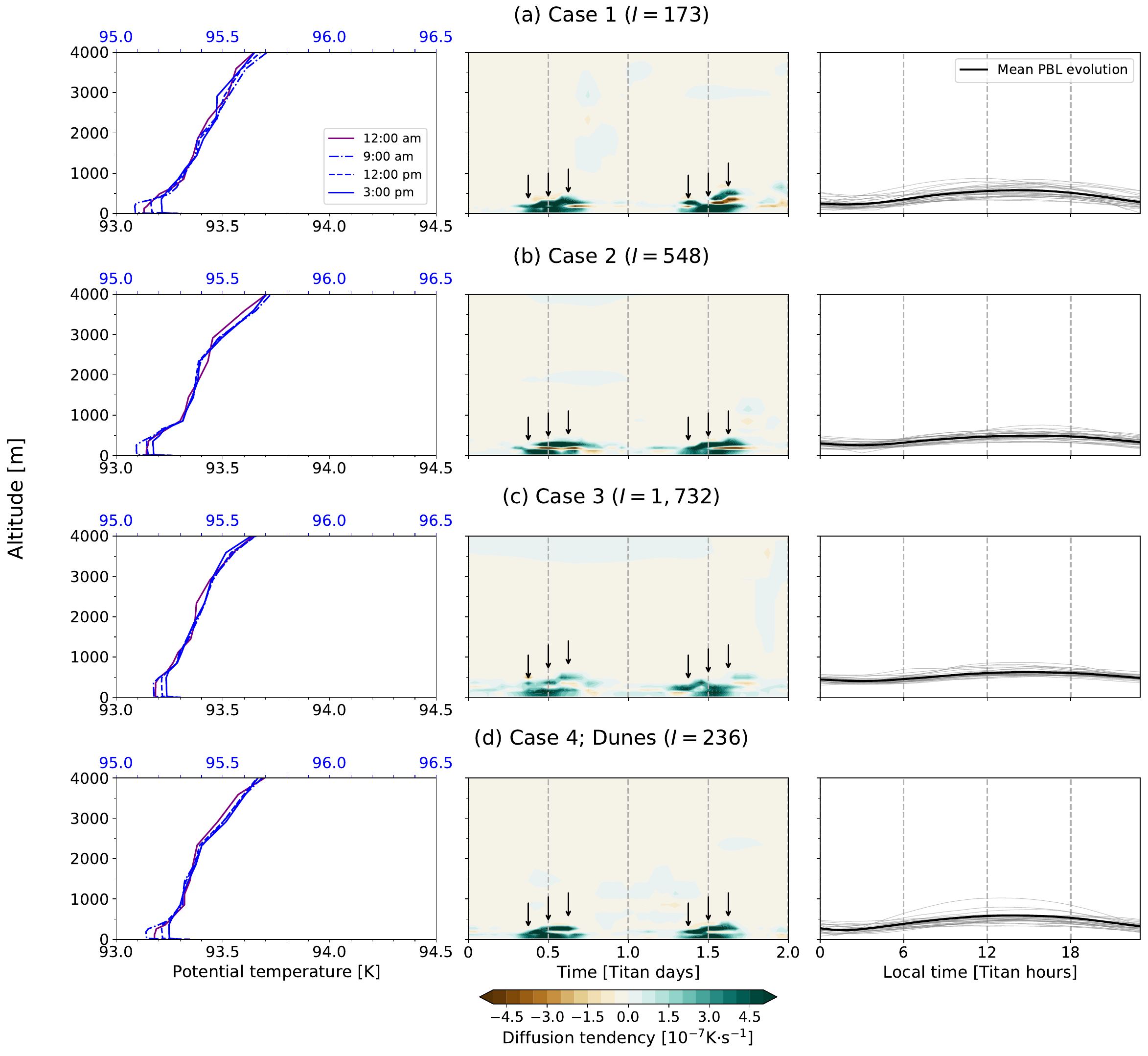}
\caption{Similar to Figure~\ref{fig:diurnal_1d_huygens_PBL}, but for the model grid point nearest the planned Dragonfly landing site, rather than the Huygens landing site.}
\label{fig:diurnal_1d_dragonfly_PBL}
\end{figure}

Figure~\ref{fig:diurnal_1d_dragonfly_PBL} shows the diurnal PBL structures at the Dragonfly landing site. Due to the reduced surface energy input and lower temperatures during this winter season—compared to the midsummer conditions at the Huygens landing site—the diagnosed PBL is shallower, with maxima of $600\pm160$~m, $610\pm180$~m, $640\pm90$~m, and $600\pm220$~m across the four cases, in contrast to 1,100-1,400~m at the Huygens site. At night, the baseline minimum PBL depths are $170\pm70$~m, $230\pm60$~m, $390\pm40$~m, and $200\pm90$~m for Cases 1–4, respectively. This results in much-reduced diurnal amplitudes (minimum to maximum) of 300–400 m, compared with 800–1,100 m at the Huygens site.

\begin{figure}[htb!]
\centering
\includegraphics[width=1.0\textwidth]{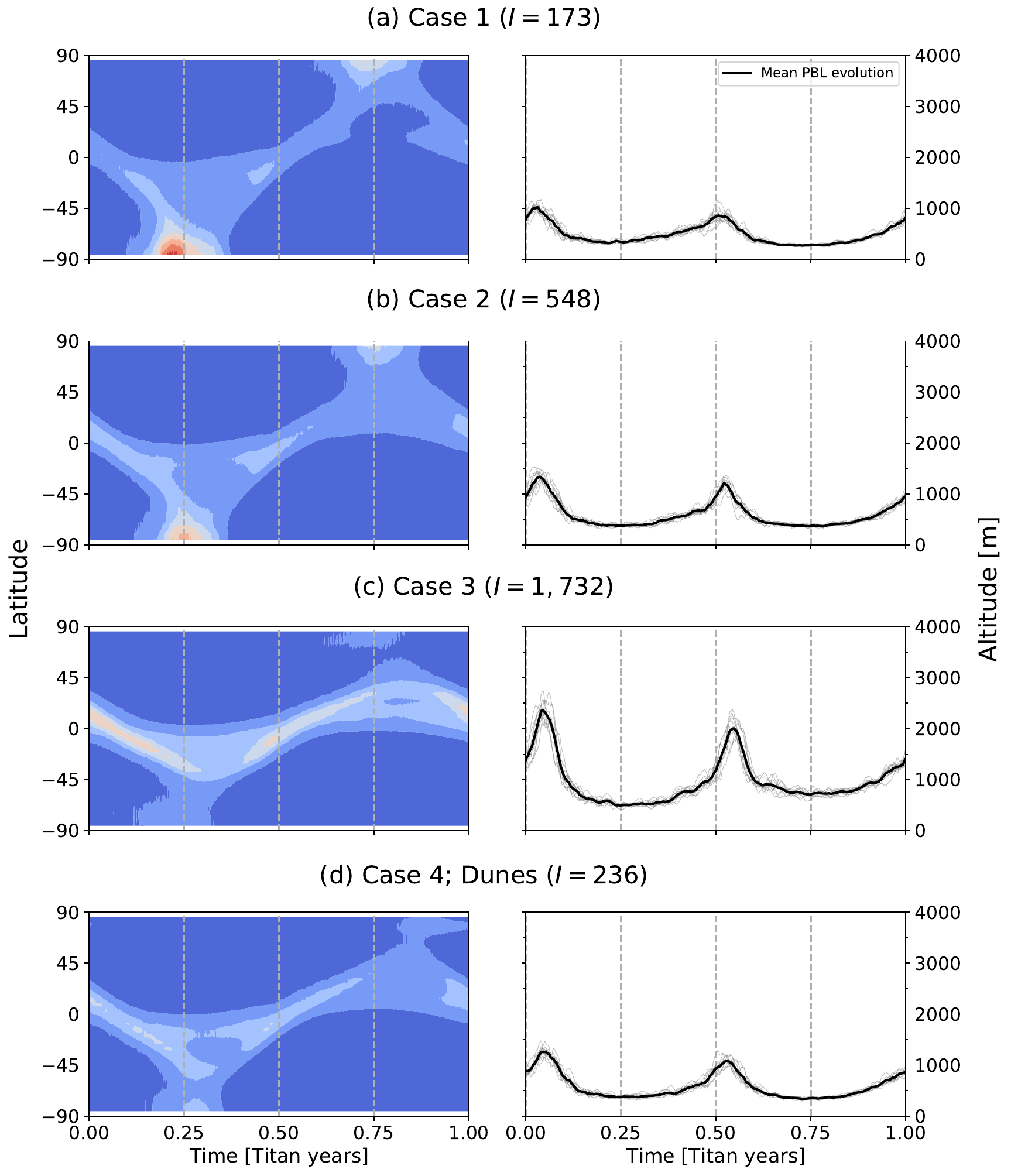}
\caption{Similar to Figure~\ref{fig:seasonal_PBL}, but for the model grid point nearest the planned Dragonfly landing site, rather than the Huygens landing site.}
\label{fig:seasonal_PBL_dragonfly}
\end{figure}

Figure~\ref{fig:seasonal_PBL_dragonfly} shows the seasonal evolution of PBL depth at the Dragonfly landing site. As at the Huygens landing site, higher thermal inertia surfaces produce deeper PBLs, with maxima of $1{,}020$, $1{,}340$, $2{,}370$, and $1{,}260$ m and minima of $270$, $360$, $490$, and $340$ m for Cases 1–4, respectively. However, a key difference is observed in the seasonal symmetry. At the Dragonfly site, the PBL depth exhibits a more symmetric seasonal pattern, owing to its near-equatorial location compared to that of Huygens: maxima occur near the equinoxes ($t \approx 0.0$ and $t \approx 0.5$), and minima near the solstices ($t \approx 0.25$ and $t \approx 0.75$). In contrast, the Huygens site displays an asymmetric pattern: a pronounced minimum at the southern winter solstice ($t \approx 0.75$), with a higher but still suppressed depth at the southern summer solstice ($t \approx 0.25$), relative to the equinoctial maxima. This asymmetry is attributed to Saturn’s orbital eccentricity, as illustrated in Figure~\ref{fig:seasonal_temp} and the left panels of Figure~\ref{fig:seasonal_PBL}.

\begin{figure}[htb!]
\centering
\includegraphics[width=0.75\textwidth]{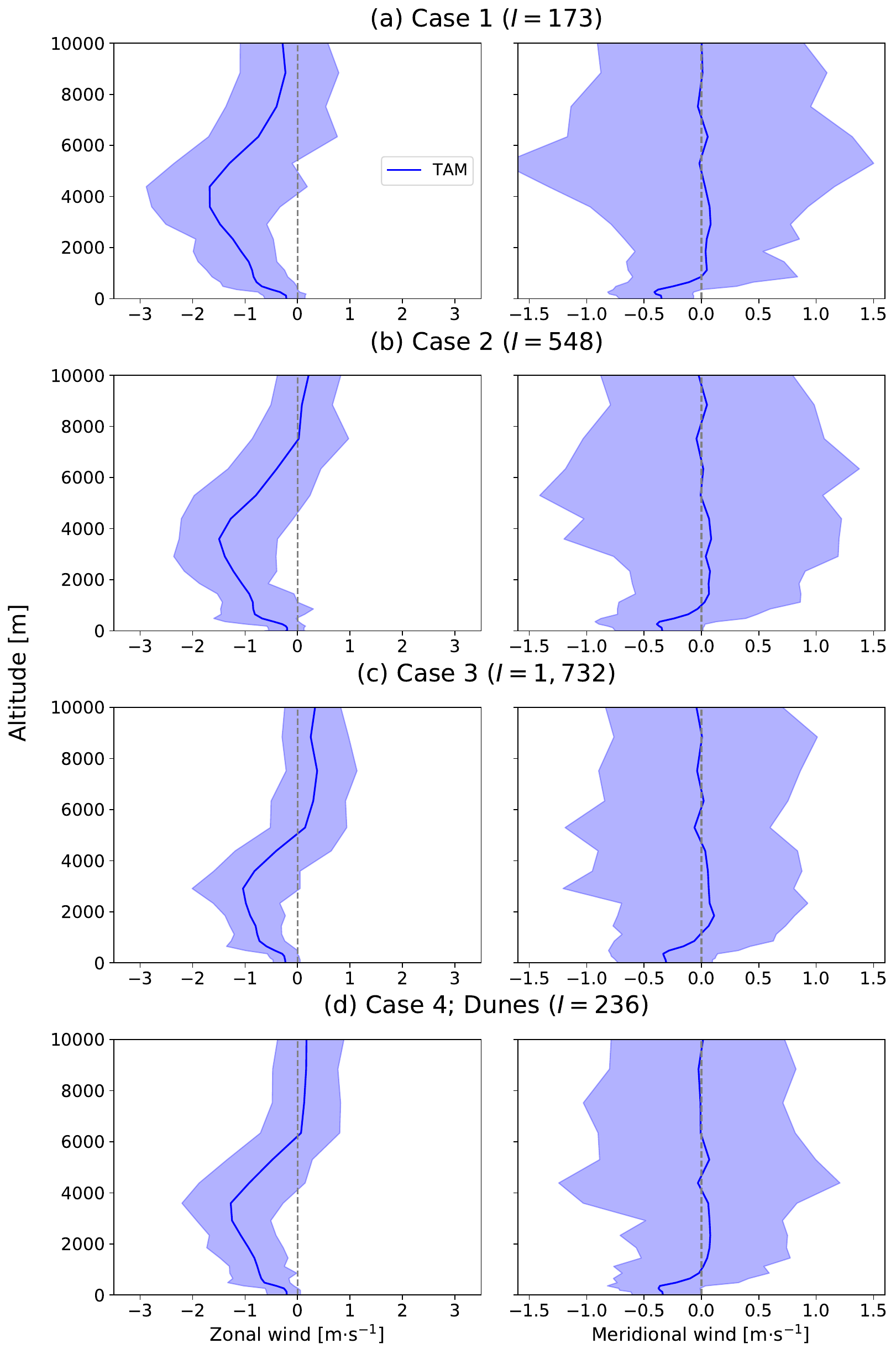}
\caption{Similar to Figure~\ref{fig:seasonal_1d_huygens_winds}, but for the model grid point nearest the planned Dragonfly landing site, rather than the Huygens landing site.}
\label{fig:seasonal_1d_dragonfly_winds}
\end{figure}

Finally, Figure~\ref{fig:seasonal_1d_dragonfly_winds} shows the seasonal wind structures at the Dragonfly landing site. Overall, the wind structures remain broadly similar between the Huygens and Dragonfly sites: westward zonal winds dominate the lowest altitudes, with southward flow near the surface and northward flow at higher altitudes. This similarity is expected, as both sites are located near Titan’s equator and are influenced by the large-scale atmospheric circulation. However, subtle differences are apparent. For instance, near-surface zonal winds at the Dragonfly site are more inclined toward negative values compared to those at the Huygens site, consistent with previous results \citep{Lora19}. The stronger westward near-surface winds at the Dragonfly site, compared to the Huygens site, may be related to its hemispheric position relative to Titan’s seasonally shifting Hadley circulation.


\section{Discussion and Conclusions}

This study presents a comprehensive reassessment of surface temperature variations on Titan across diurnal and seasonal timescales, and how these variations influence the evolution of the PBL. By combining theoretical analysis with dry GCM simulations, we identify distinct physical mechanisms that govern surface temperature variations at different timescales: subsurface heat conduction dominates the damping of diurnal variations, while atmospheric damping plays a more prominent role at seasonal timescales.

Surface temperature amplitudes are highly sensitive to thermal inertia at diurnal timescales. Low thermal inertia surfaces respond inefficiently to subsurface heat conduction, leading to larger temperature variations and stronger sensible heat fluxes. In contrast, high thermal inertia surfaces store and conduct heat more effectively, resulting in reduced temperature variation amplitude. Theoretical predictions developed in this work accurately reproduce the simulated diurnal temperature variations across different cases, confirming that subsurface heat conduction is the primary damping mechanism at short timescales. At seasonal timescales, surface temperature variations show much weaker sensitivity to thermal inertia. Only when thermal inertia is uniformly elevated (e.g., Case~3) does seasonal temperature amplitude decline significantly. This is because the product $I \cdot\sqrt{\omega}$, which characterizes the impact of subsurface heat conduction, is small compared to the atmospheric damping coefficient $a_0$ except under globally high thermal inertia. Consequently, seasonal surface temperature variations are mainly regulated by direct surface–atmosphere energy exchange through sensible heat flux, rather than by subsurface conduction.

Thermal inertia controls on the PBL are clearly evident in the simulations. At diurnal timescales, the PBL undergoes a regular cycle of growth and decay—rising from a few hundred meters during the nighttime to over 1,000 meters in the afternoon—driven by daily variations in surface heating and cooling. Surfaces with lower thermal inertia exhibit larger diurnal fluctuations in PBL depth, accompanied by stronger surface temperature variability. Notably, despite stronger daytime sensible heat fluxes, lower thermal inertia surfaces have shallower daytime PBL depths. At night, the PBL is consistently shallower over lower thermal inertia surfaces due to rapid surface cooling. These features are all linked to nighttime temperature inversions, which inhibit the development of deeper boundary layers. At seasonal timescales, the PBL responds to broader variations in insolation. As with the diurnal cycle, PBL depth increases with thermal inertia, reaching maximum depths of over 2,000 m in the highest thermal inertia case. This is consistent with previous dry GCM simulations \citep{Charnay12}, dune spacing analyses \citep{Lorenz10a}, and the dry adiabatic lapse rate inferred from Voyager IRIS observations \citep{Lindal83}. The influence of high thermal inertia on PBL structure is also evident in simulations incorporating a global thermal inertia map, where enhanced PBL depths are found over regions such as Xanadu.

Simulated winds on Titan exhibit minimal diurnal variability, which is probably linked with the small diurnal temperature swings compared to Earth. The GCM accurately captures seasonal wind structures in the lowest few kilometers, including the key zonal and meridional features observed by the Huygens probe. The simulations support the interpretation by \citet{Tokano07} that Titan’s wind profiles are shaped primarily by global-scale circulation patterns, not Ekman dynamics as on Earth. A notable seasonal transition in the meridional wind is observed in the simulations: during southern summer, the flow is southward near the surface and northward aloft (Figure~\ref{fig:seasonal_1d_huygens_winds}), while during southern winter, this pattern is reversed, with northward flow near the surface and southward aloft (not shown). This transition is characteristic of Titan’s cross-equatorial Hadley circulation.

At the Dragonfly landing site, simulations during northern midwinter show a shallower diurnal PBL—ranging from about 200 to 600 meters—due to reduced insolation compared to the midsummer conditions at the Huygens site. Nonetheless, temperature structures remain similar, and large-scale wind patterns are preserved, reflecting the shared equatorial latitude and the dominance of global atmospheric dynamics. These results suggest that Dragonfly will likely experience steady boundary layer conditions and modest winds, comparable to those encountered by Huygens.

A previous study estimated the maximum winds (i.e., the worst-case scenario that Dragonfly might encounter) associated with dry convective vortices using the equation $v_{\text{max}} = \frac{T(z=0) - T_{\text{atm}}}{2} \sqrt{\frac{2C_a}{T_{\text{atm}}}}$, obtaining a value of 2.7~m$\cdot$s$^{-1}$ with $T(z=0) - T_{\text{atm}} \approx 1.2$~K \citep{Lorenz21b}. This estimate was based on a simplified energy balance approach in which the atmospheric temperature, $T_{\text{atm}}$, is held constant, similar to that of \cite{Williams12}. In our simulations, which explicitly account for the full energy balance, applying Eq.~(7) with $|V| = 0.1$~m$\cdot$s$^{-1}$ as in \citet{Lorenz21b} yields $a_0 \approx 0.5$~(W$\cdot$m$^{-2}$$\cdot$K$^{-1}$). Combined with a sensible heat flux ranging from 0.5 to 1.0~W$\cdot$m$^{-2}$ as shown in Figure~\ref{fig:diurnal_1d_dragonfly_EB}, this results in $T(z=0) - T_{\text{atm}} \approx 1.0$ to $2.0$~K and consequently $v_{\text{max}} \approx 2.3$ to $4.6$~m$\cdot$s$^{-1}$, which is in good agreement with \citet{Lorenz21b}.

While our findings offer valuable insights, a key limitation of this study is the exclusion of methane moisture processes, which are likely to play a significant role in Titan’s climate system. Observations from the Huygens probe indicated that the surface was damp at the time of landing \citep{Zarnecki05, Lorenz06}, and our dry simulations overestimate surface temperatures by approximately 2~K. Furthermore, in several simulation cases, seasonal maxima in surface temperature and PBL depth occur at high latitudes, contrary to observational evidence indicating that Titan’s warmest regions are consistently located within 30° of the equator, as shown by Voyager IRIS data \citep{Flasar81, Flasar98} and 13 years of Cassini CIRS observations \citep{Jennings16, Jennings19}. Previous studies demonstrate that the inclusion of moist processes leads to improved latitudinal profiles of surface temperature and more realistic zonal wind structures, thereby bringing simulated climate patterns into closer agreement with observations \citep{Lora15,ML16,Turtle18,Lora19}.

This highlights the importance of including latent heat processes from methane evaporation and condensation. Prior studies have shown these to play a critical role in regulating Titan’s surface energy balance and atmospheric structure \citep{Griffith08, Mitchell09, LA17}. Incorporating methane-driven moist processes into the surface energy balance term $Q_{\text{net}}$ in Eq.~(1) will be essential for future modeling efforts. Moreover, Huygens measurements of nearly uniform methane mole fractions in the lowest 4–5~km \citep{Niemann05, Niemann10, Gautier24} suggest vigorous vertical mixing, possibly linked to the same large-scale circulation that shapes meridional wind reversals. All of this further motivates the inclusion of humidity and moist effects in future Titan PBL simulations.

In summary, this study establishes a physically grounded framework for understanding Titan’s surface temperature and boundary layer variability across diurnal and seasonal timescales. By integrating theoretical energy balance analysis with dry GCM simulations, we demonstrate the important roles of subsurface conduction and atmospheric damping in modulating surface conditions and PBL structures. While future work must incorporate methane-related processes to capture Titan’s full climate complexity, our results provide critical constraints for interpreting Huygens-era observations and offer quantitative projections for environmental conditions at the Dragonfly landing site, supporting mission planning and future data analysis.

\begin{acknowledgments}

This work was funded by NASA CDAP grant 80NSSC25K7446. The authors thank Ralph Lorenz and an anonymous reviewer for their constructive comments. SH acknowledges Xuhui Lee for valuable insights on boundary layer processes, and thanks Scot Rafkin, Victoria Hartwick and Alejandro Soto for helpful discussions. JML also acknowledges support from the Dragonfly mission. The use of the Claude AI assistant is acknowledged for grammar refinement in the preparation of the text.

The in-situ vertical profiles of temperature were measured by the Huygens Atmospheric Structure Instrument (HASI; \citealt{Fulchignoni05}), with the data available from \citet{HASI24} [\href{https://doi.org/10.17189/bcm3-4061}{doi:10.17189/bcm3-4061}].
The in-situ vertical profiles of zonal wind were measured by the Doppler Wind Experiment (DWE; \citealt{Bird05}), with the data available from \citet{DWE24} [\href{https://doi.org/10.17189/0qw0-e046}{doi:10.17189/0qw0-e046}].
Meridional winds were inferred from probe motion and tilt, reconstructed using data from the Descent Imager/Spectral Radiometer (DISR; \citealt{Karkoschka16}), available from \citet{DISR24} [\href{https://doi.org/10.17189/rx1a-ek51}{doi:10.17189/rx1a-ek51}]. The TAM simulation data are archived on \citet{HAN25} [\href{https://doi.org/10.5281/zenodo.16789992}{10.5281/zenodo.16789992}].

\end{acknowledgments}



SH and JML conceptualized the research. SH performed the formal analysis and validation, wrote the manuscript, and contributed to funding acquisition. JML provided project administration and supervision, edited the manuscript, and secured funding.


\appendix


\section{Theoretical Framework for Surface Energy Balance and Temperature Variability}

We seek a solution to the heat diffusion equation:

\begin{linenomath*}
\begin{equation}
\rho c_p \frac{\partial T(z,t)}{\partial t} = \frac{\partial}{\partial z} \left( k \frac{\partial T(z,t)}{\partial z} \right)
\label{eq:A1}
\end{equation}
\end{linenomath*}

\noindent subject to the boundary conditions:

\begin{linenomath*}
\begin{equation}
\left. k \frac{\partial T(z,t)}{\partial z} \right|_{z=0} = Q_0\exp(i \omega t) - a (T(0,t) - T_{\mathrm{atm}})
\label{eq:A2}
\end{equation}
\end{linenomath*}

\noindent and

\begin{linenomath*}
\begin{equation}
T(z \gg \delta)= \overline{T}.
\label{eq:A3}
\end{equation}
\end{linenomath*}

\noindent We consider a solution of the form:

\begin{linenomath*}
\begin{equation}
T(z, t) = \overline{T} + \tilde{T}(z) \exp(i \omega t)
\label{eq:A4}
\end{equation}
\end{linenomath*}

\noindent where $\tilde{T}(z)$ represents the complex amplitude of the temperature variation. Substituting this into \eqref{eq:A1} and assuming $k$ is constant (homogeneous subsurface) gives:  

\begin{linenomath*}
\begin{equation}
\frac{\partial^2 \tilde{T}(z)}{\partial z^2} - \alpha^2 \tilde{T}(z) = 0,
\label{eq:A8}
\end{equation}
\end{linenomath*}

\noindent where $\alpha^2 = \frac{\rho c_p i \omega}{k}$. The general solution to this second-order differential equation is $\tilde{T}(z) = A \exp({-\alpha z}) + B \exp({\alpha z})$. Since the boundary condition~\eqref{eq:A3} requires the solution to remain finite as $z \gg \delta$, the coefficient $B$ must be zero, thus $\tilde{T}(z) = A \exp({-\alpha z})$. Using $T(z, t) = \overline{T} + A \exp({-\alpha z}) \exp({i \omega t})$, we find:

\begin{linenomath*}
\begin{equation}
\frac{\partial T(z,t)}{\partial z} = \frac{\partial}{\partial z} \left( A \exp({-\alpha z}) \exp(i \omega t) \right) = -\alpha A \exp(-\alpha z) \exp(i \omega t).
\label{eq:A9}
\end{equation}
\end{linenomath*}

\noindent At $z=0$, and using the boundary condition~\eqref{eq:A2}, we obtain

\begin{linenomath*}
\begin{equation}
A = \frac{Q_0 - a\left(\overline{T} - T_{\text{atm}}\right)\exp({-i \omega t})}{k \alpha + a}.
\label{eq:A12}
\end{equation}
\end{linenomath*}

\noindent Noting that $\exp(-\alpha z)=\exp(-z/\delta)\exp(-iz/\delta)$, the subsurface temperature at any time $t$ and depth $z$ can be expressed as:

\begin{linenomath*}
\begin{equation}
T(z, t) = \overline{T} - \frac{a\left(\overline{T} - T_{\text{atm}}\right)\exp(-iz/\delta)\exp(-z/\delta)}{\sqrt{\rho c_p k} \sqrt{\omega} \exp({i \pi / 4}) + a}+
\frac{Q_0\exp({i (\omega t-z/\delta)})\exp(-z/\delta)}{\sqrt{\rho c_p k} \sqrt{\omega} \exp({i \pi / 4}) + a}.
\label{eq:A13}
\end{equation}
\end{linenomath*}

\noindent The first and second terms on the right-hand side collectively represent the mean temperature, while the third term denotes the oscillatory component. The factor $\exp(-z/\delta)$ in the third term describes the exponential decay of the thermal wave amplitude with depth, whereas $\exp(-iz/\delta)$ represents the progressive phase lag between the surface forcing and the thermal response at depth $z$. These terms capture the principal characteristics of heat diffusion in the subsurface medium. The temperature at $z=0$ becomes:

\begin{linenomath*}
\begin{equation}
T(0, t) = \overline{T} - \frac{a\left(\overline{T} - T_{\text{atm}}\right)}{\sqrt{\rho c_p k} \sqrt{\omega} \exp({i \pi / 4}) + a} 
+ \frac{Q_0}{\sqrt{\rho c_p k} \sqrt{\omega} \exp({i \pi / 4}) + a} \exp({i \omega t}).
\label{eq:A14}
\end{equation}
\end{linenomath*}

\noindent Without the atmospheric coupling coefficient $a$, Eq.~\eqref{eq:A14} simplifies to the classical solution for surface temperature variations under idealized single-harmonic periodic forcing \citep{Stull88, Garratt92}:

\begin{linenomath*}
\begin{equation}
T(0, t) = \overline{T} + \frac{Q_0}{\sqrt{\rho c_p k} \sqrt{\omega}} \exp({i( \omega t - \pi / 4)}).
\label{eq:A15}
\end{equation}
\end{linenomath*}

\noindent This solution reveals a phase lag of $\pi/4$ between the forcing and surface temperature variations when subsurface heat conduction dominates the surface temperature response. Given that the atmosphere and surface are thermally coupled through radiative and convective processes, this phase lag must also manifest in the atmospheric response. Consequently, we can express the atmospheric coupling coefficient as $a = a_0\exp(i\pi/4)$. Substituting this expression into Eq.~\eqref{eq:A14} yields:

\begin{linenomath*}
\begin{equation}
T(0, t) = \overline{T} - \frac{a_0\left(\overline{T} - T_{\text{atm}}\right)}{\sqrt{\rho c_p k} \sqrt{\omega} + a_0} + \frac{Q_0}{\sqrt{\rho c_p k} \sqrt{\omega} + a_0} \exp({i (\omega t-\pi /4)}).
\label{eq:A16}
\end{equation}
\end{linenomath*}

\noindent The amplitude of surface temperature oscillations is given by:

\begin{linenomath*}
\begin{equation}
\Delta T(z=0) \approx \frac{2Q_0}{\sqrt{\rho c_p k} \sqrt{\omega} + a_{0}} \approx \frac{\Delta Q}{\sqrt{\rho c_p k} \sqrt{\omega} + a_{0}}.
\label{eq:A17}
\end{equation}
\end{linenomath*}

\noindent where $\Delta Q$~(W$\cdot$m$^{-2}$) denotes the amplitude of fluctuations in $Q_{\text{net}}$. This expression is notably independent of the specific value of the deep boundary condition $\overline{T}$, ensuring its general applicability.










\clearpage
\bibliography{Titan}{}

\begin{thebibliography}{}
\expandafter\ifx\csname natexlab\endcsname\relax\def\natexlab#1{#1}\fi
\providecommand{\url}[1]{\href{#1}{#1}}
\providecommand{\dodoi}[1]{doi:~\href{http://doi.org/#1}{\nolinkurl{#1}}}
\providecommand{\doeprint}[1]{\href{http://ascl.net/#1}{\nolinkurl{http://ascl.net/#1}}}
\providecommand{\doarXiv}[1]{\href{https://arxiv.org/abs/#1}{\nolinkurl{https://arxiv.org/abs/#1}}}

\bibitem[{Aharonson {et~al.}(2014)Aharonson, Hayes, Hayne, Lopes, Lucas, \& Perron}]{Aharonson14}
Aharonson, O., Hayes, A.~G., Hayne, P.~O., {et~al.} 2014, in Titan: Interior, surface, atmosphere, and space environment, ed. I.~M\"{u}ller-Wodarg, C.~A. Griffith, E.~Lellouch, \& T.~E. Cravens (Cambridge: Cambridge University Press), 63--101

\bibitem[{Barnes {et~al.}(2021)Barnes, Turtle, Trainer, Lorenz, MacKenzie, Brinckerhoff, Cable, Ernst, Freissinet, Hand, Hayes, H\"{o}rst, Johnson, Karkoschka, Lawrence, Gall, Lora, McKay, Miller, Murchie, Neish, Newman, N{\'{u}}{\~{n}}ez, Panning, Parsons, Peplowski, Quick, Radebaugh, Rafkin, Shiraishi, Soderblom, Sotzen, Stickle, Stofan, Szopa, Tokano, Wagner, Wilson, Yingst, Zacny, \& St\"{a}hler}]{Barnes21}
Barnes, J.~W., Turtle, E.~P., Trainer, M.~G., {et~al.} 2021, Planet. Sci. J., 2, 130

\bibitem[{Battalio {et~al.}(2022)Battalio, Lora, Rafkin, \& Soto}]{Battalio22}
Battalio, J.~M., Lora, J.~M., Rafkin, S., \& Soto, A. 2022, Icarus, 373, 114623

\bibitem[{Bird {et~al.}(2005)Bird, Allison, Asmar, Atkinson, Avruch, Dutta-Roy, Dzierma, Edenhofer, Folkner, Gurvits, Johnston, Plettemeier, Pogrebenko, Preston, \& Tyler}]{Bird05}
Bird, M.~K., Allison, M., Asmar, S.~W., {et~al.} 2005, Nature, 438, 800

\bibitem[{Charnay {et~al.}(2015)Charnay, Barth, Rafkin, Narteau, Lebonnois, Rodriguez, Courrech~du Pont, \& Lucas}]{Charnay15}
Charnay, B., Barth, E., Rafkin, S., {et~al.} 2015, Nature Geosci., 8, 362

\bibitem[{Charnay \& Lebonnois(2012)}]{Charnay12}
Charnay, B., \& Lebonnois, S. 2012, Nature Geosci., 5, 106

\bibitem[{Courtin \& Kim(2002)}]{Courtin02}
Courtin, R., \& Kim, S.~J. 2002, Planet. Space Sci., 50, 309

\bibitem[{Dickinson(1988)}]{Dickinson88}
Dickinson, R.~E. 1988, J.\ Climate, 1, 1086

\bibitem[{Faulk {et~al.}(2020)Faulk, Lora, Mitchell, \& Milly}]{Faulk20}
Faulk, S.~P., Lora, J.~M., Mitchell, J.~L., \& Milly, P. C.~D. 2020, Nature Astron., 4, 390

\bibitem[{Flasar(1998)}]{Flasar98}
Flasar, F.~M. 1998, Planet. Space Sci., 46, 1125

\bibitem[{Flasar \& Achterberg(2009)}]{Flasar09}
Flasar, F.~M., \& Achterberg, R.~K. 2009, Philos. T. R. Soc. A, 367, 649

\bibitem[{Flasar {et~al.}(2014)Flasar, Achterberg, \& Schinder}]{Flasar14}
Flasar, F.~M., Achterberg, R.~K., \& Schinder, P.~J. 2014, in Titan: Interior, surface, atmosphere, and space environment, ed. I.~M\"{u}ller-Wodarg, C.~A. Griffith, E.~Lellouch, \& T.~E. Cravens (Cambridge: Cambridge University Press), 102--121

\bibitem[{Flasar {et~al.}(1981)Flasar, Samuelson, \& Conrath}]{Flasar81}
Flasar, F.~M., Samuelson, R.~E., \& Conrath, B.~J. 1981, Nature, 292, 693

\bibitem[{Folkner {et~al.}(2024)Folkner, Bird, Preston, \& Asmar}]{DWE24}
Folkner, W., Bird, M., Preston, R., \& Asmar, S. 2024, HP-SSA-DWE-2-3-DESCENT-V1.0: Huygens Probe Doppler Wind Experiment V1.0,  NASA Planetary Data System, \dodoi{10.17189/0qw0-e046}

\bibitem[{{Fulchignoni} {et~al.}(2005){Fulchignoni}, {Ferri}, {Angrilli}, {Ball}, {Bar-Nun}, {Barucci}, {Bettanini}, {Bianchini}, {Borucki}, {Colombatti}, {Coradini}, {Coustenis}, {Debei}, {Falkner}, {Fanti}, {Flamini}, {Gaborit}, {Grard}, {Hamelin}, {Harri}, {Hathi}, {Jernej}, {Leese}, {Lehto}, {Lion Stoppato}, {L{\'o}pez-Moreno}, {M{\"a}kinen}, {McDonnell}, {McKay}, {Molina-Cuberos}, {Neubauer}, {Pirronello}, {Rodrigo}, {Saggin}, {Schwingenschuh}, {Seiff}, {Sim{\~o}es}, {Svedhem}, {Tokano}, {Towner}, {Trautner}, {Withers}, \& {Zarnecki}}]{Fulchignoni05}
{Fulchignoni}, M., {Ferri}, F., {Angrilli}, F., {et~al.} 2005, Nature, 438, 785

\bibitem[{Garratt(1992)}]{Garratt92}
Garratt, J.~R. 1992, The {A}tmospheric {B}oundary {L}ayer (New York: Cambridge University Press)

\bibitem[{Gautier {et~al.}(2024)Gautier, Serigano, Das, Coutelier, Hörst, Szopa1, Vinatier, \& Trainer}]{Gautier24}
Gautier, T., Serigano, J., Das, K., {et~al.} 2024, A. \& A., 690, A165

\bibitem[{Griffith {et~al.}(2008)Griffith, McKay, \& Ferri}]{Griffith08}
Griffith, C.~A., McKay, C.~P., \& Ferri, F. 2008, Astrophys. J., 687, L41

\bibitem[{Han(2025)}]{HAN25}
Han, S. 2025, {T}itan {A}tmospheric {M}odel ({TAM}) dry {GCM} simulations for Titan's planetary boundary layer (supporting "{T}hermal {I}nertia {C}ontrols on {T}itan's {S}urface {T}emperature and {P}lanetary {B}oundary {L}ayer {S}tructure"),  Zenodo, \dodoi{10.5281/zenodo.16789992}

\bibitem[{{HASI Team}(2024)}]{HASI24}
{HASI Team}. 2024, HP-SSA-HASI-2-3-4-MISSION-V1.1: Huygens HASI Raw and Calibrated Data Archive V1.1,  NASA Planetary Data System, \dodoi{10.17189/bcm3-4061}

\bibitem[{Hayes(2016)}]{Hayes16}
Hayes, A.~G. 2016, Annu. Rev. Earth Planet. Sci., 44, 57

\bibitem[{Hinson {et~al.}(2010)Hinson, Pätzold, Tellmann, Häusler, \& Tyler}]{Hinson08}
Hinson, D.~P., Pätzold, M., Tellmann, S., Häusler, B., \& Tyler, G.~L. 2010, Icarus, 198, 57

\bibitem[{Holtslag \& Boville(1993)}]{Holtslag93}
Holtslag, A.~A., \& Boville, B.~A. 1993, J.\ Climate, 6, 1825

\bibitem[{Huber(2024)}]{DISR24}
Huber, L. 2024, Descent Imager/Spectral Radiometer (DISR) for HP Bundle,  NASA Planetary Data System, \dodoi{10.17189/rx1a-ek51}

\bibitem[{Jennings {et~al.}(2016)Jennings, Cottini, Nixon, Achterberg, Flasar, Kunde, Romani, Samuelson, Mamoutkine, Gorius, Coustenis, \& Tokano}]{Jennings16}
Jennings, D.~E., Cottini, V., Nixon, C.~A., {et~al.} 2016, Astrophys. J. Lett., 816, L17

\bibitem[{Jennings {et~al.}(2019)Jennings, Tokano, Cottini, Nixon, Achterberg, Flasar, Kunde, Romani, Samuelson, Segura, Gorius, Guandique, Kaelberer, \& Coustenis}]{Jennings19}
Jennings, D.~E., Tokano, T., Cottini, V., {et~al.} 2019, Astrophys. J. Lett., 877, L8

\bibitem[{Karkoschka(2016)}]{Karkoschka16}
Karkoschka, E. 2016, Icarus, 270, 326

\bibitem[{Lebonnois {et~al.}(2012)Lebonnois, Burgalat, Rannou, \& Charnay}]{Lebonnois12}
Lebonnois, S., Burgalat, J., Rannou, P., \& Charnay, B. 2012, Icarus, 218, 707

\bibitem[{Lebonnois {et~al.}(2018)Lebonnois, Schubert, F., \& Spiga}]{Lebonnois18}
Lebonnois, S., Schubert, G., F., F., \& Spiga, A. 2018, Icarus, 314, 149

\bibitem[{Lee(2016)}]{Lee16}
Lee, X. 2016, Springer

\bibitem[{{Lewis} {et~al.}(2023){Lewis}, {Lombardo}, {Read}, \& Lora}]{Lewis23}
{Lewis}, N.~T., {Lombardo}, N.~A., {Read}, P.~L., \& Lora, J.~M. 2023, Planet. Sci. J., 4, 149

\bibitem[{{Lindal} {et~al.}(1983){Lindal}, {Wood}, {Hotz}, {Sweetnam}, {Eshleman}, \& {Tyler}}]{Lindal83}
{Lindal}, G.~F., {Wood}, G.~E., {Hotz}, H.~B., {et~al.} 1983, Icarus, 53, 348

\bibitem[{{Lombardo} \& Lora(2023)}]{Lombardo23b}
{Lombardo}, N.~A., \& Lora, J.~M. 2023, J. Geophys. Res. Planets, 128, e2023JE008061

\bibitem[{Lora(2024)}]{Lora24}
Lora, J.~M. 2024, Icarus, 422, 116241

\bibitem[{Lora \& {\'A}d{\'a}mkovics(2017)}]{LA17}
Lora, J.~M., \& {\'A}d{\'a}mkovics, M. 2017, Icarus, 286, 270

\bibitem[{Lora {et~al.}(2022)Lora, Battalio, Yap, \& Baciocco}]{Lora22}
Lora, J.~M., Battalio, J.~M., Yap, M., \& Baciocco, C. 2022, Icarus, 384, 115095

\bibitem[{Lora {et~al.}(2015)Lora, Lunine, \& Russell}]{Lora15}
Lora, J.~M., Lunine, J.~I., \& Russell, J.~L. 2015, Icarus, 250, 516

\bibitem[{Lora {et~al.}(2019)Lora, Tokano, Vatant~d'Ollone, Lebonnois, \& Lorenz}]{Lora19}
Lora, J.~M., Tokano, T., Vatant~d'Ollone, J., Lebonnois, S., \& Lorenz, R.~D. 2019, Icarus, 333, 113

\bibitem[{Lora {et~al.}(2025)Lora, Turtle, \& Mitchell}]{Lora25}
Lora, J.~M., Turtle, E.~P., \& Mitchell, J.~L. 2025, in {Titan After Cassini-Huygens}, ed. R.~M.~C. Lopes, C.~Elachi, I.~C.~F. M\"uller-Wodarg, \& A.~Solomonidou (Elsevier), 201--237

\bibitem[{Lorenz(2021)}]{Lorenz21b}
Lorenz, R.~D. 2021, Icarus, 354, 114062

\bibitem[{Lorenz {et~al.}(2010)Lorenz, Claudin, Andreotti, Radebaugh, \& Tokano}]{Lorenz10a}
Lorenz, R.~D., Claudin, P., Andreotti, B., Radebaugh, J., \& Tokano, T. 2010, Icarus, 205, 719

\bibitem[{Lorenz \& Radebaugh(2009)}]{Lorenz09}
Lorenz, R.~D., \& Radebaugh, J. 2009, Geophys. Res. Lett., 36, L03202

\bibitem[{Lorenz {et~al.}(2006)Lorenz, {Wall}, {Radebaugh}, {Boubin}, {Reffet}, {Janssen}, {Stofan}, {Lopes}, {Kirk}, {Elachi}, {Lunine}, {Mitchell}, {Paganelli}, {Soderblom}, {Wood}, {Wye}, {Zebker}, {Anderson}, {Ostro}, {Allison}, {Boehmer}, {Callahan}, {Encrenaz}, {Ori}, {Francescetti}, {Gim}, {Hamilton}, {Hensley}, {Johnson}, {Kelleher}, {Muhleman}, {Picardi}, {Posa}, {Roth}, {Seu}, {Shaffer}, {Stiles}, {Vetrella}, {Flamini}, \& {West}}]{Lorenz06}
Lorenz, R.~D., {Wall}, S., {Radebaugh}, J., {et~al.} 2006, Science, 312, 724

\bibitem[{Lorenz {et~al.}(2021)Lorenz, MacKenzie, Neish, Le~Gall, Turtle, Barnes, Trainer, Werynski, Hedgepeth, \& Karkoschka}]{Lorenz21a}
Lorenz, R.~D., MacKenzie, S.~M., Neish, C.~D., {et~al.} 2021, Planet. Sci. J., 2, 24

\bibitem[{MacKenzie {et~al.}(2019)MacKenzie, Lora, \& Lorenz}]{Mackenzie19b}
MacKenzie, S.~M., Lora, J.~M., \& Lorenz, R.~D. 2019, J. Geophys. Res. Planets, 124, 1728

\bibitem[{McKay {et~al.}(1989)McKay, Pollack, \& Courtin}]{McKay89}
McKay, C.~P., Pollack, J.~B., \& Courtin, R. 1989, Icarus, 80, 23

\bibitem[{McKay {et~al.}(1991)McKay, Pollack, \& Courtin}]{Mckay91}
---. 1991, Science, 253, 1118

\bibitem[{Mellor \& Yamada(1982)}]{Mellor82}
Mellor, G., \& Yamada, T. 1982, Rev. Geophys., 20, 851

\bibitem[{Mitchell \& Lora(2016)}]{ML16}
Mitchell, J.~L., \& Lora, J.~M. 2016, Annu. Rev. Earth Planet. Sci., 44, 353

\bibitem[{Mitchell {et~al.}(2009)Mitchell, Pierrehumbert, Frierson, \& Caballero}]{Mitchell09}
Mitchell, J.~L., Pierrehumbert, R.~T., Frierson, D. M.~W., \& Caballero, R. 2009, Icarus, 203, 250

\bibitem[{{Niemann} {et~al.}(2005){Niemann}, {Atreya}, {Bauer}, {Carignan}, {Demick}, {Frost}, {Gautier}, {Haberman}, {Harpold}, {Hunten}, {Israel}, {Lunine}, {Kasprzak}, {Owen}, {Paulkovich}, {Raulin}, {Raaen}, \& {Way}}]{Niemann05}
{Niemann}, H.~B., {Atreya}, S.~K., {Bauer}, S.~J., {et~al.} 2005, Nature, 438, 779

\bibitem[{Niemann {et~al.}(2010)Niemann, Atreya, Demick, Gautier, Haberman, Harpold, Kasprzak, Lunine, Owen, \& Raulin}]{Niemann10}
Niemann, H.~B., Atreya, S.~K., Demick, J.~E., {et~al.} 2010, J. Geophys. Res., 115, E12006

\bibitem[{Petrosyan {et~al.}(2011)Petrosyan, Galperin, Larsen, Lewis, Määttänen, Read, Renno, Rogberg, Savijärvi, Siili, Spiga, Toigo, \& Vázquez}]{Petrosyan11}
Petrosyan, A., Galperin, B., Larsen, S.~E., {et~al.} 2011, Rev. Geophys., 49, RG3005

\bibitem[{Price(1977)}]{Price77}
Price, J.~C. 1977, J. Geophys. Res., 82, 2582

\bibitem[{Putzig {et~al.}(2005)Putzig, Mellon, Kretke, \& Arvidson}]{Putzig05}
Putzig, N.~E., Mellon, M.~T., Kretke, K.~A., \& Arvidson, R.~E. 2005, Icarus, 173, 325

\bibitem[{Samuelson {et~al.}(1997)Samuelson, Nath, \& Borysow}]{Samuelson97}
Samuelson, R.~E., Nath, N.~R., \& Borysow, A. 1997, Planet. Space Sci., 45, 959

\bibitem[{Schinder {et~al.}(2020)Schinder, Flasar, Marouf, French, Anabtawi, Barbinis, Fleischman, \& Achterberg}]{Schinder20}
Schinder, P.~J., Flasar, F.~M., Marouf, E.~A., {et~al.} 2020, Icarus, 345, 113720

\bibitem[{Schinder {et~al.}(2011)Schinder, Flasar, Marouf, French, McGhee, Kliore, Rappaport, Barbinis, Fleischman, \& Anabtawi}]{Schinder11}
---. 2011, Icarus, 215, 460

\bibitem[{Schinder {et~al.}(2012)Schinder, Flasar, Marouf, French, McGhee, Kliore, Rappaport, Barbinis, Fleischman, \& Anabtawi}]{Schinder12}
---. 2012, Icarus, 221, 1020

\bibitem[{Spiga {et~al.}(2010)Spiga, Forget, Lewis, \& Hinson}]{Spiga10}
Spiga, A., Forget, F., Lewis, S., \& Hinson, D.~P. 2010, Q. J. R. Meteorol. Soc., 136, 414

\bibitem[{Stull(1988)}]{Stull88}
Stull, R.~B. 1988, Springer

\bibitem[{Tokano(2005)}]{Tokano05}
Tokano, T. 2005, Icarus, 173, 222

\bibitem[{Tokano(2007)}]{Tokano07}
---. 2007, Planet. Space Sci., 55, 1990

\bibitem[{Tokano {et~al.}(2006)Tokano, Ferri, Colombatti, M{\"a}kinen, \& Fulchignoni}]{Tokano06b}
Tokano, T., Ferri, F., Colombatti, G., M{\"a}kinen, T., \& Fulchignoni, M. 2006, J. Geophys. Res. Planet., 111, E08007

\bibitem[{Tomasko {et~al.}(2008)Tomasko, B\'ezard, Doose, Engel, Karkoschka, \& Vinatier}]{Tomasko08c}
Tomasko, M.~G., B\'ezard, B., Doose, L., {et~al.} 2008, Planet. Space Sci., 56, 648

\bibitem[{Turtle {et~al.}(2018)Turtle, Perry, Barbara, Del~Genio, Rodriguez, Sotin, Lora, Faulk, Corlies, Kelland, MacKenzie, West, McEwen, Lunine, Pitesky, Ray, \& Roy}]{Turtle18}
Turtle, E.~P., Perry, J.~E., Barbara, J.~M., {et~al.} 2018, Geophys. Res. Lett., 45, 5320

\bibitem[{Williams {et~al.}(2012)Williams, McKay, \& Persson}]{Williams12}
Williams, K.~E., McKay, C.~P., \& Persson, F. 2012, Planet. Space Sci., 60, 376

\bibitem[{Zarnecki {et~al.}(2005)Zarnecki, Leese, Hathi, Ball, Hagermann, Towner, Lorenz, McDonnell, Green, Patel, Ringrose, Rosenberg, Atkinson, Paton, Banaszkiewicz, Clark, Ferri, Fulchignoni, Ghafoor, Kargl, Svedhem, Delderfield, Grande, Parker, Challenor, \& Geake}]{Zarnecki05}
Zarnecki, J.~C., Leese, M.~R., Hathi, B., {et~al.} 2005, Nature, 438, 792

\end{thebibliography}
\bibliographystyle{aasjournal}



\begin{table}
\caption{Four simulation cases investigated in this study, including their prescribed thermal properties and derived parameters. Cases 1–3 use uniform thermal properties with increasing thermal conductivity, while Case 4 adopts a spatially varying thermal inertia map (convective lakes scenario) based on terrain types from \citet{Mackenzie19b}.}
\begin{tabular}{l c c c c c c}  
\hline
  & $^{(a)}\rho c_p$ & $^{(b)}k$ & $^{(c)}I = \sqrt{\rho c_p k}$ & $^{(d)}\delta_{a} = \sqrt{\frac{2k}{\omega_{a} \rho c_p}}$ & $^{(e)}\delta_{d} = \sqrt{\frac{2k}{\omega_{d} \rho c_p}}$ & \\  
\hline
 Case 1 & $3 \times 10^{6}$ & 0.01 & 173 & 1.0 & 0.04 &\\   
 Case 2 & $3 \times 10^{6}$ & 0.1 & 548 & 3.1 & 0.1 &\\
 Case 3 & $3 \times 10^{6}$ & 1.0 & 1,732 & 9.9 & 0.4 &\\
 \multicolumn{6}{l}{Case 4: $^{(f)}$A spatially varying thermal inertia map is adopted.} \\
 
\hline
\multicolumn{7}{l}{$^{(a)}$J$\cdot$m$^{-3}\cdot$K$^{-1}$, $^{(b)}$W$\cdot$m$^{-1}\cdot$K$^{-1}$, $^{(c)}$J$\cdot$m$^{-2}\cdot$s$^{-0.5}\cdot$K$^{-1}$, $^{(d)(e)}$m} \\ 
\multicolumn{7}{l}{$^{(d)}$annual thermal skin depth corresponding to $\omega_{a}=\frac{2\pi}{29.5 \times 365 \times 86400}$ s$^{-1}$} \\
\multicolumn{7}{l}{$^{(e)}$diurnal thermal skin depth corresponding to $\omega_{d}=\frac{2\pi}{16 \times 86400}$ s$^{-1}$} \\
\multicolumn{7}{l}{$^{(f)}$\cite{Mackenzie19b}} \\
\hline
\label{tab:table1}
\end{tabular}
\end{table}

\end{document}